%% file: master.tex
\documentclass[nojss,article]{jss}
\usepackage{bm} 
\usepackage{hyperref}
\usepackage{amsmath}
\usepackage{natbib}
\usepackage{nicefrac}
\usepackage{tabularx}
\usepackage{graphicx}
\usepackage{subfig}
\usepackage{url}
\usepackage{subfloat}
\makeatletter
\newcommand{\distas}[1]{\mathbin{\overset{#1}{\kern\z@\sim}}}%
\newsavebox{\mybox}\newsavebox{\mysim}
\newcommand{\distras}[1]{%
  \savebox{\mybox}{\hbox{\kern3pt$\scriptstyle#1$\kern3pt}}%
  \savebox{\mysim}{\hbox{$\sim$}}%
  \mathbin{\overset{#1}{\kern\z@\resizebox{\wd\mybox}{\ht\mysim}{$\sim$}}}%
}
\makeatother
\usepackage[utf8,latin1]{inputenc} 

\newcommand{\BetaBin}{\operatorname{BetaBin}}

\newcommand{\Var}{\operatorname{Var}}

\newcommand{\logit}{\operatorname{logit}}
\newcommand{\NB}{\operatorname{NB}}

\author{Ma{\"e}lle Salmon\\Robert Koch Institute \And 
        Dirk Schumacher\\Robert Koch Institute \And
        Michael H{\"o}hle\\ Stockholm University,\\Robert Koch Institute }
\title{Monitoring Count Time Series in R: Aberration Detection in Public Health Surveillance}
\Plainauthor{Ma\"elle Salmon, Dirk Schumacher, Michael H\"ohle} 
\Shorttitle{\pkg{surveillance}: Aberration detection in R} 

\Abstract{
 \input{abstract}

}

\Keywords{R, \texttt{surveillance}, outbreak detection, statistical process control}
\Plainkeywords{R, surveillance, outbreak detection, statistical process control} 

\Address{
  Ma\"{e}lle Salmon\\
 Department for Infectious Diseases Epidemiology\\
Robert Koch Institut Berlin\\
Seestrasse 10\\
13353 Berlin, Germany\\

  E-mail: \email{SalmonM@rki.de}\\
  URL: \url{http://www.normalesup.org/~masalmon/}\\
  \smallskip\\
    Dirk Schumacher\\
 Department for Infectious Diseases Epidemiology\\
Robert Koch Institut Berlin\\
Seestrasse 10\\
13353 Berlin, Germany\\

  E-mail: \email{SchumacherD@rki.de}\\
  URL: \url{http://www.dirk-schumacher.net/}\\
    \smallskip\\
    Michael H\"{o}hle\\
Department of Mathematics\\
Stockholm University\\
Kr\"{a}ftriket\\
106 91 Stockholm,Sweden

  E-mail: \email{hoehle@math.su.se}\\
  URL: \url{http://www.math.su.se/~hoehle/}
}

\begin{document}

\maketitle

\section*{Introduction}
  \addcontentsline{toc}{section}{Introduction}
\input{introduction}

\section{Getting to know the basics of the package}
\input{part1}
\section[Using surveillance in selected contexts]{Using \pkg{surveillance} in selected contexts}
\label{sec:using}
\input{part2}
\section[Implementing surveillance in routine monitoring]{Implementing \pkg{surveillance} in 
routine monitoring}
\label{sec:routine}
\input{part3}
\section*{Discussion}
\addcontentsline{toc}{section}{Discussion and outlook}
\input{discussion}

\section*{Acknowledgements}
\input{acknowledgements}
\bibliographystyle{jss}
\bibliography{biblio}
\end{document}

%% file: abstract.tex
Public health surveillance aims at lessening disease burden, \textit{e.g.}, in case of infectious diseases by timely recognizing emerging outbreaks. Seen from a statistical perspective, this implies the use of appropriate methods for monitoring time series of aggregated case reports. This paper presents the tools for such automatic aberration detection offered by the \textsf{R} package \pkg{surveillance}. We introduce the functionality for the visualization, modelling and monitoring of surveillance time series.
With respect to modelling we focus on univariate time series modelling based on generalized linear models (GLMs), multivariate GLMs, generalized additive models and generalized additive models for location, shape and scale. This ranges from illustrating implementational improvements and extensions of the well-known Farrington algorithm, \textit{e.g.}, by spline-modelling or by treating it in a Bayesian context. Furthermore, we look at categorical time series and address overdispersion using beta-binomial or Dirichlet-Multinomial modelling. With respect to monitoring we consider detectors based on either a Shewhart-like single timepoint comparison between the observed count and the predictive distribution or by likelihood-ratio based cumulative sum methods. Finally, we illustrate how \pkg{surveillance} can support aberration detection in practice by integrating it into the monitoring workflow of a public health institution. Altogether, the present article shows how well \pkg{surveillance} can support automatic aberration detection in a public health surveillance context.

%% file: introduction.tex
Nowadays, the fight against infectious diseases does not only require treating patients and setting up measures for prevention but also demands the timely recognition of emerging outbreaks in order to avoid their expansion. Along these lines, health institutions such as hospitals and public health authorities collect and store information about health events -- typically represented as individual case reports containing clinical information, and subject to specific case definitions. Analysing these data is crucial. It enables situational awareness in general and in particular the timely detection of aberrant counts empowering the prevention of additional disease cases through early interventions. For any specific aggregation of characteristics of events, such as over-the-counter sales of pain medication, new cases of foot-and-mouth disease among cattle, or adults becoming sick with hepatitis C in Germany, data can be represented as time series of counts with days, weeks, months or years as time units of the aggregation. Abnormally high or low values at a given time point can reveal critical issues such as an outbreak of the disease or a malfunction of data transmission. Thus, identifying aberrations in the collected data is decisive, for human as well as for animal health. \\
\\ In this paper we present the \proglang{R} package \pkg{surveillance}, which implements a range of methods for aberration detection in time series of counts and proportions. Statistical algorithms provide an objective and reproducible analysis of the data and allow the automation of time-consuming aspects of the monitoring process. In the recent years, a variety of such tools has flourished in the literature. Reviews of methods for aberration detection in time series of counts can be found in~\citet{Buckeridge2005}~and~\citet{Unkel2012}. However, the great variety of statistical algorithms for aberration detection can be a hurdle to practitioners wishing to find a suitable method for their data. It is our experience that ready-to-use and understandable implementation and the possibility to use the methods in a routine and automatic fashion are the criteria most important to the epidemiologists.\\
\\ The package offers an open-source implementation of state-of-the-art methods for the prospective detection of outbreaks in count data time series with established methods, as well as the visualization of the analysed time series. With the package, the practitioner can introduce statistical surveillance into routine practice without too much difficulty. As far as we know, the package is now used in several public health institutions in Europe: at the National Public Health Institute of Finland, at the Swedish Institute for Communicable Disease Control, at the French National Reference Centre for Salmonella, and at the Robert Koch Institute (RKI) in Berlin. The use of \pkg{surveillance} at the RKI shall be the focus of this paper. The package also provides many other functions serving epidemic modelling purposes. Such susceptible-infectious-recovered based models and their extensions towards regression based approaches are documented in other works~\citep{held_etal2005,held_etal2006,meyer_etal2012,meyer.etal2014}. \\
\\ The present paper is designed as an extension of two previous articles about the \pkg{surveillance} package published as~\citet{hoehle2007} and~\citet{hoehle_mazick2009}. On the one hand, the paper aims at giving an overview of the new features added to the package since the publication of the two former papers. On the other hand it intends to illustrate how well the \pkg{surveillance} package can support routine practical disease surveillance by presenting the current surveillance system of infectious diseases at the RKI. \\
\\ This paper is structured as follows. Section 1 gives an introduction to the data structure used in the package for representing and visualizing univariate or multivariate time series. Furthermore, the structure and use of aberration detection algorithms are explained. Section 2 leads the reader through different surveillance methods available in the package. Section 3 describes the integration of such methods in a complete surveillance system as currently in use at the RKI. Finally, a discussion rounds off the work.

%% file: part1.tex
\DefineVerbatimEnvironment{Sinput}{Verbatim}{fontshape=sl}
\DefineVerbatimEnvironment{SinputSmall}{Verbatim}{fontshape=sl}
\DefineVerbatimEnvironment{Soutput}{Verbatim}{fontshape=s1}
\DefineVerbatimEnvironment{Scode}{Verbatim}{fontshape=sl}

The 
package provides a central S4 data class \code{sts} to capture multivariate or univariate time series. All further methods use objects of this class as an input. 
Therefore we first describe how to use the \code{sts} class and then,
 as all monitoring methods of the package conform to the same syntax, a typical call of a function for aberration detection will be presented. Furthermore, the visualization of time series and of the results of their monitoring is depicted.
\subsection{How to store time series and related information}

\indent\indent In \pkg{surveillance}, time series of counts and related information are encoded in a specific S4-class called \code{sts} (\textit{surveillance time series}) that represents 
possibly multivariate time series of counts. Denote the counts as $\left( y_{it} ; i = 1, \ldots,m, t = 1,  \ldots, n \right)$,
where $n$ is the length of the time series and $m$ is the number of entities, \textit{e.g.}, geographical regions, hospitals or age groups,
being
monitored. An example which we shall look at in more details is a time series representing the weekly counts of cases of infection with \textit{Salmonella Newport} in all 16 federal states of Germany 
from 2004 to 2013 with $n=525$ weeks and $m=16$ geographical units. Infections with \textit{Salmonella Newport}, a subtype of \textit{Salmonella}, can trigger gastroenteritis, prompting
the seek of medical care. Infections with \textit{Salmonella} are notifiable in Germany since 2001 with data being forwarded to the RKI by federal states health authorities on behalf of the local health authorities.
 
\subsubsection*{Slots of the class \code{sts}}
The key slots of the \code{sts} class are those describing the observed counts and the corresponding time periods of the aggregation. The observed counts $\left(y_{it}\right)$ are stored in the $n \times m$ matrix  \code{observed}.
A number of other slots characterize time. First, \code{epoch} denotes the corresponding time period of the aggregation. If the Boolean \code{epochAsDate} is \code{TRUE}, 
\code{epoch} is the numeric representation of \code{Date} objects corresponding to each observation in \code{observed}. If the Boolean \code{epochAsDate} is \code{FALSE}, 
\code{epoch} is the time index $1 \leq t \leq n$ of each of these observations. Then, \code{freq} is the number of observations per year: 365 for
daily data, 52 for weekly data and 12 for monthly data. 
Finally, \code{start} is a vector representing the origin of the time series with two values that are the year and the epoch within
that year for the first observation of the time series -- \code{c(2014,1)} for a time series starting on the first week of 2014 for instance.

Other slots enable the storage of additional information. Known aberrations are recorded in the Boolean slot \code{state} of the same dimensions as \code{observed}
 with \code{TRUE} indicating an outbreak and \code{FALSE} indicating the absence of any known aberration. The monitored population in each of the units is stored in 
 slot \code{populationFrac}, which gives either proportions or numbers. The geography of the zone under surveillance is accessible through slot \code{map} 
 which is an object of class \code{SpatialPolygonsDataFrame}~\citep{sp1,sp2} providing a shape of the $m$ areas which are monitored
and slot \code{neighbourhood}, which is a symmetric matrix of Booleans size $m^2$ stating the neighbourhood
matrix. Slot \code{map} is pertinent when units are geographical units, whereas \code{neighbourhood} could be useful in any case, \textit{e.g.}, for storing a contact matrix between age groups for modelling purposes. 
 Finally, if monitoring has been performed on the data the information on its control arguments and its results are stored in \code{control}, \code{upperbound} and \code{alarm} presented in Section~\ref{sec:howto}.
\subsubsection*{Creation of an object of class \code{sts}}
The creation of a \code{sts} object is straightforward, requiring a call to the function \code{new} together with the slots to be assigned as arguments. The input of data from external files is one possibility for getting the counts 
as is described in \citet{hoehle_mazick2009}.
To exemplify the process we shall use weekly counts of \textit{Salmonella Newport} in Germany loaded using \code{data("salmNewport")}. "). Alternatively, one can use coercion methods to convert between the \texttt{ts} class and the \texttt{sts} class. Note that this only converts the content of the slot \texttt{observed}, that is,

\begin{Schunk}
\begin{Sinput}
R> all.equal(observed(salmNewport),observed(as(as(salmNewport,"ts"),"sts")))
\end{Sinput}
\end{Schunk}

Using the \texttt{ts} class as intermediate step also allows the conversion between other time series classes, e.g. from packages \pkg{zoo}~\citep{zoo} or \pkg{xts}~\citep{xts}.

 \setkeys{Gin}{height=7cm, width=15cm}
\begin{figure}
\begin{center}  
\includegraphics{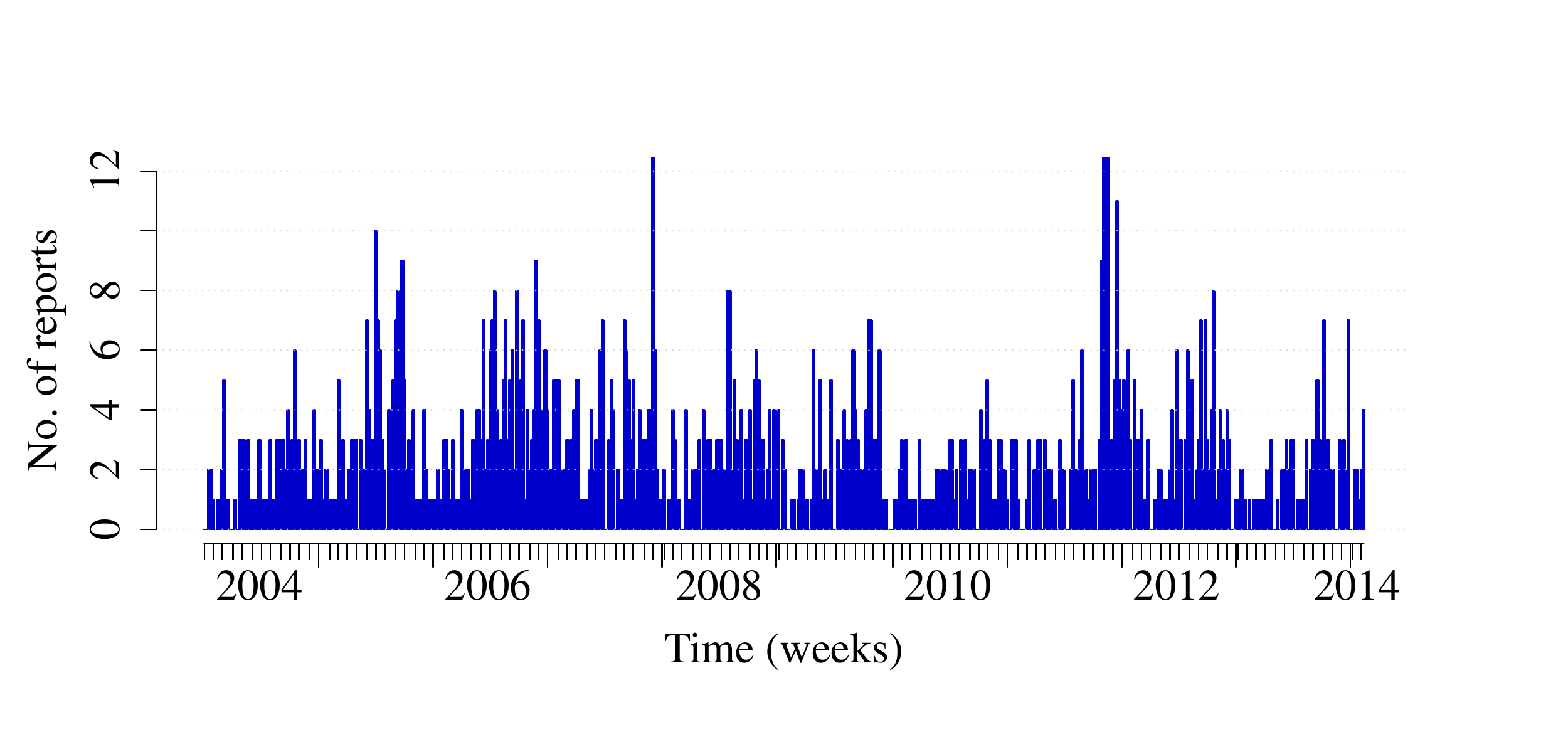}
\end{center}
\vspace{-1cm}
\caption{Weekly number of cases of S. Newport in Germany, 2004-2013.}
\label{fig:Newport}
\end{figure}
\subsubsection*{Basic manipulation of objects of the class \code{sts}}
\begin{figure}
\hspace{-1em}
\subfloat[]{
 
\includegraphics[width=9cm]{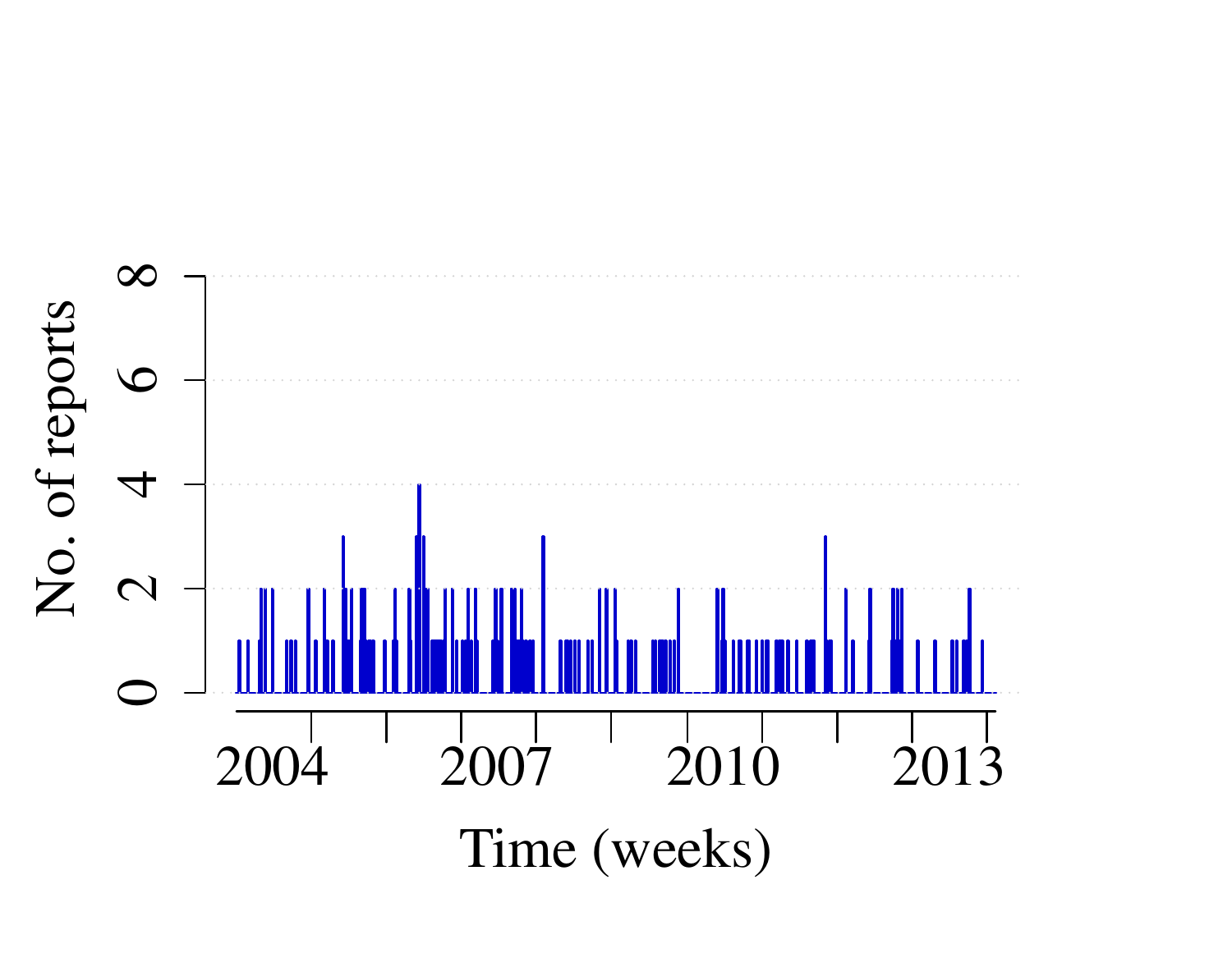}
}\hspace{-3em}%
\subfloat[]{
\includegraphics[width=9cm]{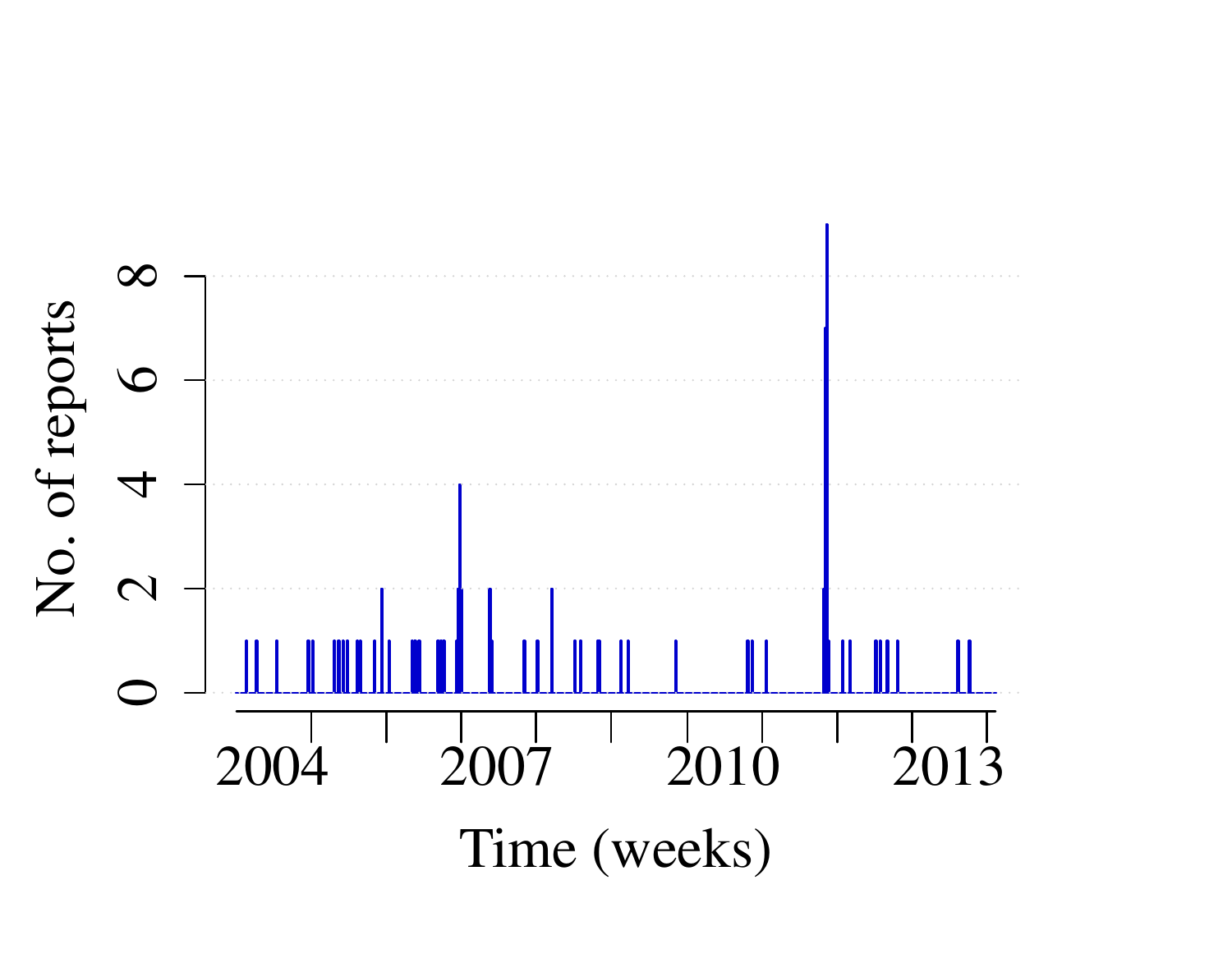}
}
\caption{Weekly count of S. Newport in the German federal states (a) Bavaria and (b) Berlin.}
\label{fig:unit}
\end{figure}

This time series above is represented as a multivariate \code{sts}-object whose dimensions correspond
to the 16 German federal states. Values are weekly counts so \code{freq=52}.
 Weeks are here handled as \code{Date} objects by setting \code{epochAsDate} to \code{TRUE}. One can thus for instance get the weekday of the date by calling \code{weekdays(salmNewport)}. Furthermore, one can use the function \code{format} (and the package specific platform independent version \code{dateFormat}) to obtain \code{strftime} compatible formatting of the epochs. Another advantage of using \code{Date} objects is that the plot functions have been re-written for better management of ticks and labelling of the x-axis based on \code{strtime} compatible conversion specifications.
For example, to get ticks at all weeks corresponding to the first week in a month as well as all weeks corresponding to the first in a year while placing labels consisting of the year at the median index per year:
\begin{Schunk}
\begin{Sinput}
R> plot(salmNewport,
 	xaxis.tickFreq = list("%V" = atChange, "%m" = atChange,
 						"%G" = atChange),
 	xaxis.labelFreq = list("%Y" = atMedian),
 	xaxis.labelFormat = "%Y", type = observed ~ time)
\end{Sinput}
\end{Schunk}
which is shown in Fig.~\ref{fig:Newport}. Here, the \code{atChange} and \code{atMedian} functions are small helper functions and the  respective tick lengths are controlled by the \pkg{surveillance} specific option \code{surveillance.options("stsTickFactors")}. Actually \code{sts}-objects
can be plotted using different options: \code{type = observed ~ time} produces the time series for whole Germany as shown on Fig.~\ref{fig:Newport}, whereas \code{type = observed ~ time | unit}
is a panelled graph with each panel representing the time series of counts of a federal state as seen on Fig.~\ref{fig:unit}. 
 \setkeys{Gin}{height=7cm, width=9cm}

Once created one can use typical subset operations on a \code{sts}-object: for instance \code{salmNewport}\\\code{[1:10,"Berlin"]} is a new \code{sts}-object 
with weekly counts for Berlin during the 10 first weeks of the initial dataset; \code{salmNewport[isoWeekYear(epoch(salmNewport))\$ISOYear<=2010,]} uses the \code{surveillance}'s \code{isoWeekYear()} 
function to get a \code{sts}-object with weekly counts for all federal states up to 2010. Moreover, one can take advantage of the \proglang{R} function \code{aggregate()}. For instance,  \code{aggregate(salmNewport,by="unit")} returns a \code{sts}-object representing weekly counts of
\textit{Salmonella Newport} in Germany as a whole, whereas \code{aggregate(salmNewport,\\by="time")} corresponds to the total count of cases in each federal state over the whole period.
\subsection{How to use aberration detection algorithms}
\label{sec:howto}
Monitoring algorithms in the package operate on objects of the class \code{sts} as described below.
\subsubsection*{Statistical framework for aberration detection}
We introduce the framework for aberration detection on an
univariate time series of counts $\left\{y_t,\> t=1,2,\ldots\right\}$. Surveillance aims
at detecting an \textit{aberration}, that is to say, an important change in the process occurring at an unknown time $\tau$. 
This change can be a step increase of the counts of cases or a more gradual change~\citep{Sonesson2003}.

Based on the possibility of such a change, for each time $t$ we want to differentiate between the two states \textit{in-control} and \textit{out-of-control}. At any timepoint $t_0\geq 1$, the available information -- \textit{i.e.}, past counts -- is defined as $\bm{y}_{t_0} = \left\{
y_t\>;\> t\leq t_0\right\}$. Detection is based on a statistic $r(\cdot)$ with resulting
 alarm time $T_A = \min\left\{ t_0\geq 1 : r(\bm{y}_{t_0}) > g\right\}$ where $g$ is a known threshold.
Functions for aberration detection thus use past data to estimate $r(\bm{y}_{t_0})$, and compare it to the threshold $g$,
above which the current count can be considered as suspicious and thus doomed as \textit{out-of-control}.

Threshold values and alarm Booleans for each timepoint of the monitored range are 
saved in the slots \code{upperbound} and \code{alarm}, of the same dimensions as \code{observed}, while the method parameters used for computing the threshold values 
and alarm Booleans 
are stored in the slot \code{control}.
\subsubsection*{Aberration detection in the package}
To perform such a monitoring of the counts of cases, one has to choose one of the surveillance algorithms of the package -- this choice will be the topic
of Section~\ref{sec:using}. Then, one must indicate which part of the time series or \code{range} has to be monitored -- for instance the current year. Lastly, one needs to specify the parameters specific to the algorithm.
\subsubsection*{Example with the EARS C1 method}
We will illustrate the basic principle by using the \code{earsC}~function~that implements the EARS (Early Aberration Detection System) methods of the CDC as described in~\citet{SIM:SIM3197}. This algorithm
is especially convenient in situations when little historic information is available. It offers three variants called C1, C2 and C3.

Here we shall expand on C1 for which the baseline are the 7 timepoints before the assessed timepoint $t_0$, that is to say
 $\left(y_{t_0-7},\ldots,y_{t_0-1}\right)$. The expected value is the mean of the baseline. The method is based on a statistic called $C_{t_0}$ defined as
$C_{t_0}= \nicefrac{(y_{t_0}-\bar{y}_{t_0})}{s_{t_0}}$, where $$\bar{y}_{t_0}= \frac{1}{7} \cdot\sum_{i=t_0-7}^{t_0-1} y_i  \textnormal{ and } s_{t_0}^2= \frac{1}{7-1} \cdot\sum_{i=t_0-7}^{t_0-1} \left(y_i - \bar{y}_{t_0}\right)^2.$$ 
Under the null hypothesis of no outbreak, it is assumed that $C_{t_0} \stackrel{H_0}{\sim} {N}(0,1)$. The upperbound $U_{t_0}$ is defined by a $(1-\alpha)\cdot 100\%$ confidence interval for the mean:
 $U_{t_0}= \bar{y}_{t_0} + z_{1-\alpha}s_{t_0}$ where  $z_{1-\alpha}$ is the $(1-\alpha)$'th quantile of the standard normal distribution. An alarm is raised if $y_{t_0} > U_{t_0}$.

 The output of the algorithm is a \code{sts}-object that contains subsets of slots \code{observed}, \code{population} and \code{state} defined by the range of timepoints specified in the input -- 
\textit{e.g} the last 20 timepoints of the timeseries,
and with the slots \code{upperbound} and \code{alarm} filled by the output of the algorithm. 
Information relative to the \code{range} of data to be monitored 
and to the parameters of the algorithm, such as \code{alpha} for \code{earsC}, has to be formulated in the slot \code{control}. 
This information is also stored in the slot \code{control} of the returned \code{sts}-object for later inspection. 
\begin{Schunk}
\begin{Sinput}
R> in2011 <- which(isoWeekYear(epoch(salmNewport))$ISOYear == 2011)
R> salmNewportGermany <- aggregate(salmNewport, by = "unit")
R> control <- list(range = in2011, method = "C1", alpha = 0.05)
R> surv <- earsC(salmNewportGermany, control = control)		  
R> plot(surv)
\end{Sinput}
\end{Schunk}
 \setkeys{Gin}{height=7cm, width=15cm}
\begin{figure}
\begin{center}  
  
\includegraphics{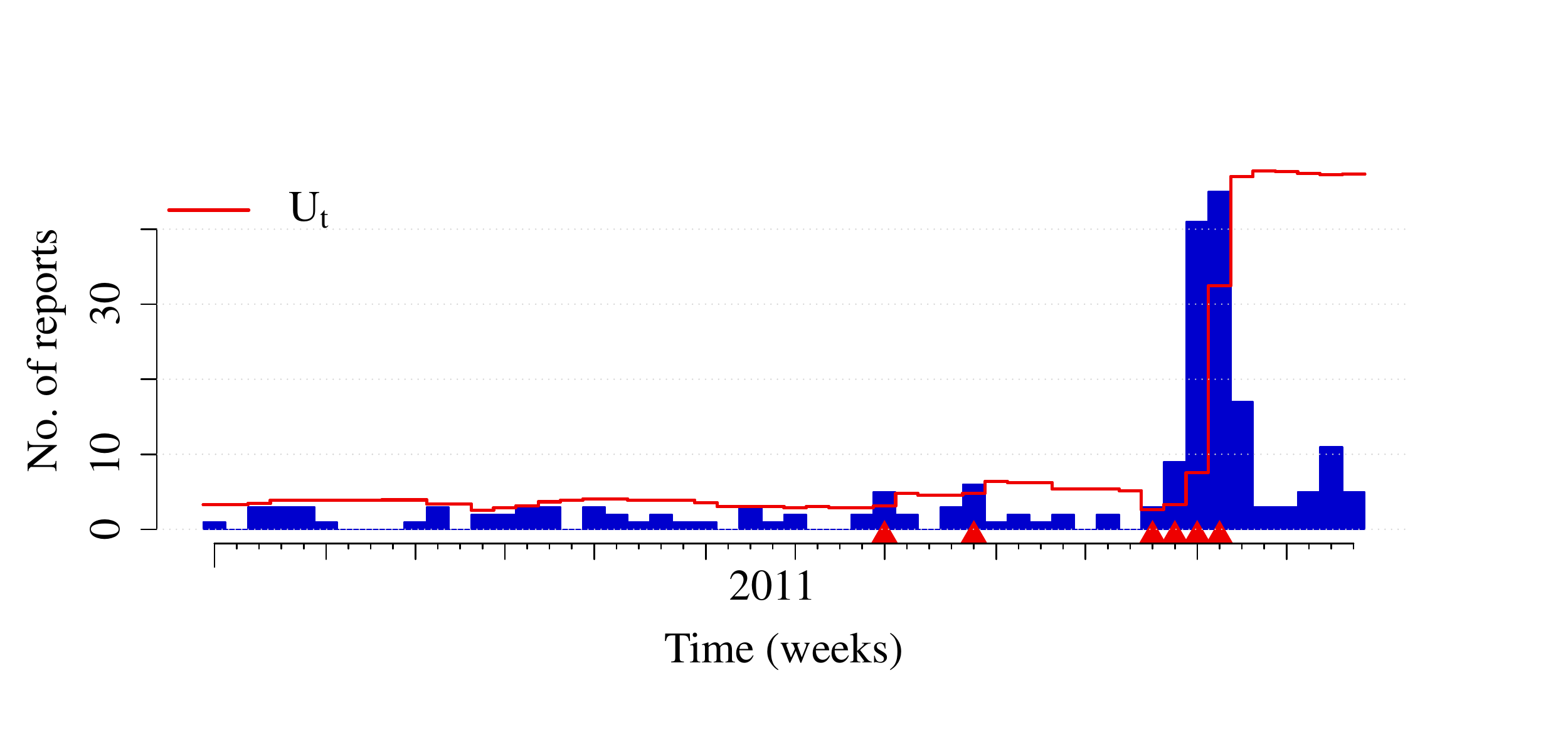}
\end{center}
\vspace{-1cm}
\caption{Weekly reports of S. Newport in Germany in 2011 monitored by the EARS C1 method. The line
represents the upperbound calculated by the algorithm. Triangles indicate alarms that are
the timepoints where the observed number of counts is higher than the upperbound.}
\label{fig:NewportEARS}
\end{figure}

The \code{sts}-object is easily visualized
using the function \code{plot} as depicted in Fig.~\ref{fig:NewportEARS}, which shows the upperbound as a dashed line and the alarms -- timepoints 
where the upperbound has been exceeded -- as triangles. The four last alarms
correspond to a known outbreak in 2011 due to sprouts~\citep{Newport2011}. One sees that the upperbound right after the outbreak
is affected by the outbreak: it is very high, so that a smaller outbreak would not be detected.

The EARS methods C1, C2 and C3 are simple in that they only use information from the very recent past. This is appropriate when data has only been collected for a short time
or when one expects the count to be fairly constant. However, data from the less recent past often encompass relevant information about \textit{e.g.}, seasonality and time trend, that one should take into
account when estimating the expected count and the associated threshold. For instance, ignoring an increasing time trend could decrease sensitivity. Inversely, overlooking
an annual surge in counts during the summer could decrease specificity.
 Therefore, it is advisable to use detection methods whose underlying models incorporate essential characteristics of time series of disease count data
such as overdispersion, seasonality, time trend and presence of past outbreaks in the records~\citep{Unkel2012,Shmueli2010}. Moreover, the EARS methods do not compute a proper prediction interval for the 
current count. Sounder statistical methods will be reviewed in the next section. 

%% file: part2.tex
\DefineVerbatimEnvironment{Sinput}{Verbatim}{fontshape=sl}
\DefineVerbatimEnvironment{SinputSmall}{Verbatim}{fontshape=sl}
\DefineVerbatimEnvironment{Soutput}{Verbatim}{fontshape=s1}
\DefineVerbatimEnvironment{Scode}{Verbatim}{fontshape=sl}
\setkeys{Gin}{width=0.9\textwidth}
More than a dozen algorithms for aberration detection are implemented in the package. Among those,
this section presents a set of representative algorithms, which are already in routine application at several public health institutions or which we think have the potential to become so.
First we describe the Farrington method introduced by~\citet{farrington96} together with the improvements proposed by~\citet{Noufaily2012}.  
As a Bayesian counterpart to these methods we present the BODA method published by~\citet{Manitz2013} which allows
the easy integration of covariates. All these methods perform one-timepoint detection in that they detect aberrations only when the count at
the currently monitored timepoint is above the threshold. Hence, no accumulation of evidence takes place. As an extension, we introduce an implementation of the negative binomial cumulative sum (CUSUM) of~\citet{hoehle_paul2008} that allows the detection
of sustained shifts by accumulating evidence over several timepoints. Finally, we present a method suitable for categorical data described in~\citet{hoehle2010} that is also based on cumulative sums.
\subsection{One size fits them all for count data}
Two implementations of the Farrington method, which is currently \textit{the} method of choice at European public health institutes~\citep{hulth_etal2010}, exist in the package. First, the original method as described in ~\citet{farrington96} is implemented as function \code{farrington}. Its use was already described in \citet{hoehle_mazick2009}.
 Now, the newly implemented function \code{farringtonFlexible} supports the use of this \textit{original method} as well as of the \textit{improved method} built on suggestions made by~\citet{Noufaily2012} for improving the specificity without reducing the sensitivity. 

 In the function \code{farringtonFlexible} one can choose to use the original method or the improved method by specification of appropriate \code{control} arguments. 
Which variant of the algorithm is to be used is determined by the contents of the \code{control} slot. 
In the example below, \code{control1} corresponds to the use of the original method and \code{control2} indicates the options for the improved method.
\begin{Schunk}
\begin{Sinput}
R> control1 <- list(range = in2011, noPeriods = 1,
                  b = 4, w = 3, weightsThreshold = 1,
                  pastWeeksNotIncluded = 3, pThresholdTrend = 0.05,
                  thresholdMethod = "delta")
R> control2 <- list(range = in2011, noPeriods = 10,
                  b = 4, w = 3, weightsThreshold = 2.58,
                  pastWeeksNotIncluded = 26, pThresholdTrend = 1,
                  thresholdMethod = "nbPlugin")
\end{Sinput}
\end{Schunk}

In both cases the steps of the algorithm are the same. In a first step, an overdispersed Poisson generalized linear model with log link is fitted to the reference data $\bm{y}_{t_0} \subseteq \left\{
y_t\>;\> t\leq t_0\right\}$, where $\E(y_t)=\mu_t$  with $\log \mu_t = \alpha + \beta t$ and $\Var(y_t)=\phi\cdot\mu_t$ and where $\phi\geq1$ is ensured.              

\begin{figure}
\begin{center}
\subfloat[$\texttt{noPeriods}=2$]{

\includegraphics[width=0.45\textwidth]{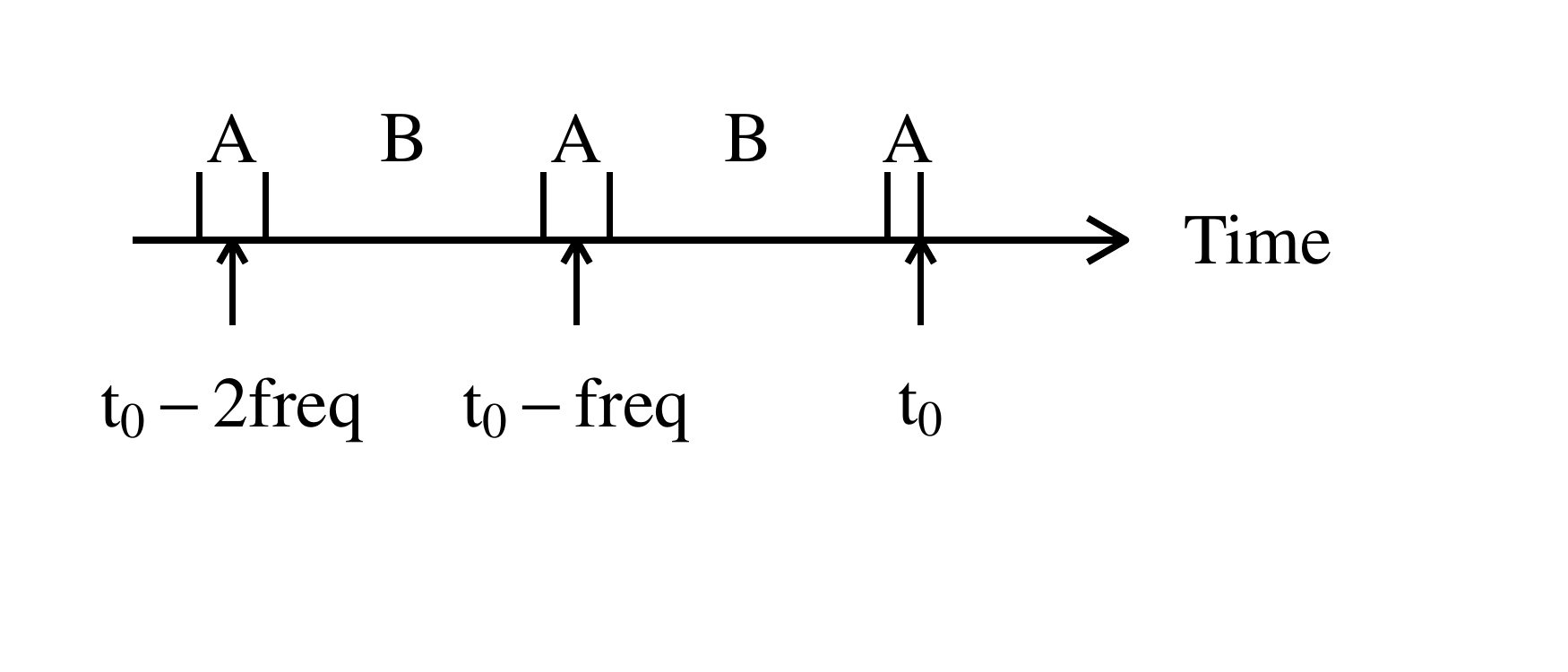}
}
\qquad
\subfloat[$\texttt{noPeriods}=3$]{
\includegraphics[width=0.45\textwidth]{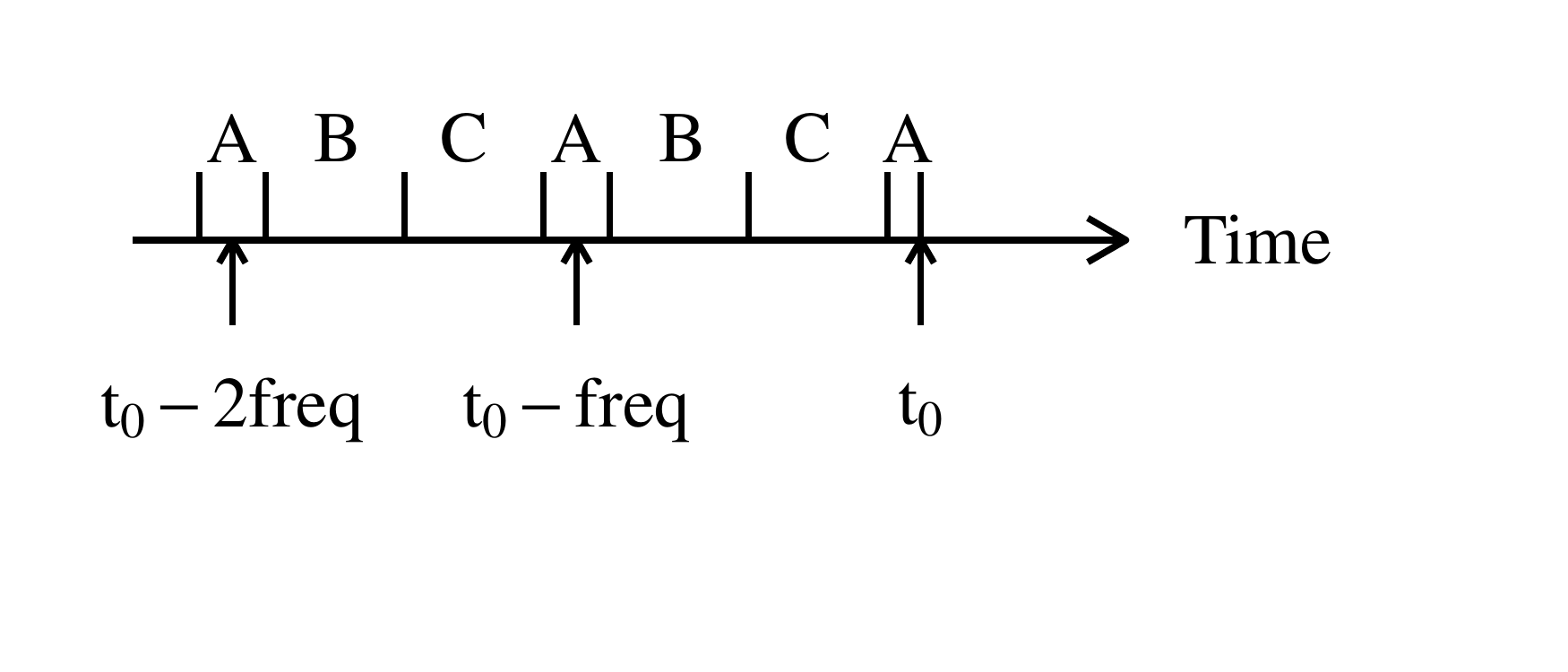}
}

\caption{Construction of the noPeriods-level factor to account for seasonality, depending on the value of the half-window size $w$ and of the freq of the 
data. Here the number of years to go back in the past $b$ is 2. 
Each level of the factor variable corresponds to a period delimited by ticks and is denoted by a character. The windows around $t_0$ are respectively of size $2w+1$,~$2w+1$ 
and $w+1$. The segments between them are divided into the other periods so that they 
have the same length up to rounding.}
\label{fig:fPlot}
\end{center}
\end{figure}

The original method took seasonality into account by using a subset of the available data as reference data for fitting the GLM: \code{w} timepoints
centred around the timepoint located $1,2,\ldots,b$ years before $t_0$, amounting to a total $b \cdot (2w+1)$ reference values.
Yet, it was shown in~\citet{Noufaily2012} that the algorithm performs better when using more historical data. In order to do do so without disregarding seasonality, the authors
introduced a zero order spline with 11 knots, which can be conveniently represented as a 10-level factor. We have extended this idea in our implementation so that one can choose an arbitrary number of periods
in each year. Thus, $\log \mu_t = \alpha + \beta t + f(t)$ where $f(t)$ is a zero order periodic spline with $\texttt{noPeriods}+1$ knots represented as a $\texttt{noPeriods}$-level factor. 
The algorithm uses \code{w}, \code{b} and \texttt{noPeriods} to deduce the length of periods so they have 
the same length up to rounding. An exception is the reference window centred around $t_0$. Figure~\ref{fig:fPlot} shows a minimal example, where each character corresponds to a different period. Note that setting $\texttt{noPeriods} = 1$ corresponds to using the original method with only a subset of the data: 
there is only one period defined per year, the reference window around $t_0$ and other timepoints are not included in the model.

 \setkeys{Gin}{height=3cm, width=7cm}
Moreover, it was shown in \citet{Noufaily2012} that is is better to exclude the last 26 weeks before $t_0$ from the baseline 
in order to avoid reducing sensitivity when an outbreak has started recently before $t_0$. In the \code{farringtonFlexible} function, one controls this by specifying \code{pastWeeksNotIncluded}, which is 
the number of last timepoints before $t_0$ that are not to be used. The default value is 26.

Lastly, in the new implementation a population offset can be included in the GLM by setting \code{populationBool} to \code{TRUE} and supplying the possibly time-varying population size in the \code{population} slot of the \code{sts}-object, but this will not be discussed further here.

In a second step, the expected number of counts $\mu_{t_0}$ is predicted for the current timepoint $t_0$ using this GLM. An upperbound $U_{t_0}$ is calculated based on this predicted value and its variance. The two versions of the algorithm make different assumptions for this calculation.
The 
original method assumes that a transformation of the prediction error $g\left(y_{t_0}-\hat{\mu}_{t_0}\right)$ is normally distributed, for instance when using the identity transformation $g(x)=x$ one obtains
$$y_{t_0} - \hat{\mu}_0 \sim \mathcal{N}(0,\Var(y_{t_0}-\hat{\mu}_0))\cdot$$
The upperbound of the prediction interval is then calculated 
based on this distribution. First we have that
$$ \Var(y_{t_0}-\hat{\mu}_{t_0}) = \Var(\hat{y}_{t_0}) + \Var(\hat{\mu}_{t_0})=\phi\mu_0+\Var(\hat{\mu}_{t_0}) $$
 with $\Var(\hat{y}_{t_0})$ being the variance of an observation and $\Var(\hat{\mu}_{t_0})$ being the variance of the estimate. The threshold, defined as the upperbound of a one-sided $(1-\alpha)\cdot 100\%$ confidence interval, is 
$$U_{t_0} = \hat{\mu}_0 + z_{1-\alpha}\widehat{\Var}(y_{t_0}-\hat{\mu}_{t_0})\cdot$$

This method can be used by setting the control option \code{thresholdMethod} equal to "\code{delta}".
 However, a weakness 
of this procedure is the normality assumption itself, so that an alternative was presented in \citet{Noufaily2012} and implemented as \code{thresholdMethod="Noufaily"}.  
The central assumption of this approach is that 
$y_{t_0} \sim \NB\left(\mu_{t_0},\nu\right)$,
with $\mu_{t_0}$ the mean of the distribution and $\nu=\nicefrac{\mu_{t_0}}{\phi-1}$ its overdispersion parameter. In this parameterization, we still have $\E(y_t)=\mu_t$ and $\Var(y_t)=\phi\cdot\mu_t$ with $\phi>1$ -- otherwise a Poisson distribution is assumed for the 
observed count. 
The threshold is defined as a quantile of the negative binomial distribution with
plug-in estimates $\hat{\mu}_{t_0}$ and $\hat{\phi}$. Note that this disregards the estimation uncertainty in $\hat{\mu}_{t_0}$ and $\hat{\phi}$. As a consequence, the method "\code{muan}" (\textit{mu} for $\mu$ and \textit{an} for asymptotic normal) tries to solve the problem by
using the asymptotic normal distribution of $(\hat{\alpha},\hat{\beta})$ to derive the upper $(1-\alpha)\cdot 100\%$ quantile of the asymptotic normal distribution of $\hat{\mu}_{t_0}=\hat{\alpha}+\hat{\beta}t_0$. Note that this does not reflect all
 estimation uncertainty because it disregards the estimation uncertainty of $\hat{\phi}$.
Note also that for time series where the variance of the estimator is large, the upperbound also ends up being very large. 
Thus, the method "\code{nbPlugin}" seems to provide information that is easier to interpret by epidemiologists but with "\code{muan}" being more statistically correct.

In a last step, the observed count $y_{t_0}$ is compared to the upperbound $U_{t_0}$ and an alarm is raised if $y_{t_0} > U_{t_0}$.
In both cases the fitting of the GLM involves three important steps. First, the algorithm performs an optional power-transformation for skewness correction and variance stabilisation, 
depending on the value of the parameter \code{powertrans} in the \code{control} slot. Then, the significance of the time trend is checked. The time trend is included only when significant at a chosen level \code{pThresholdTrend}, when
they are more than three years reference data and if no overextrapolation occurs because of the time trend. Lastly, past outbreaks are reweighted based on their Anscombe residuals. In \code{farringtonFlexible} the limit for reweighting past counts, \code{weightsThreshold},
can be specified by the user. If the Anscombe residuals of a count is higher than \code{weightsThreshold} it is reweighted accordingly in a second fitting of the GLM. \citet{farrington96} used a value of $1$
whereas \citet{Noufaily2012} advises a value of $2.56$ so that the reweighting procedure is less drastic, because it also shrinks the variance of the observations.  

\begin{figure}
\hspace{-1em}
\subfloat[]{
 
\includegraphics[width=9cm]{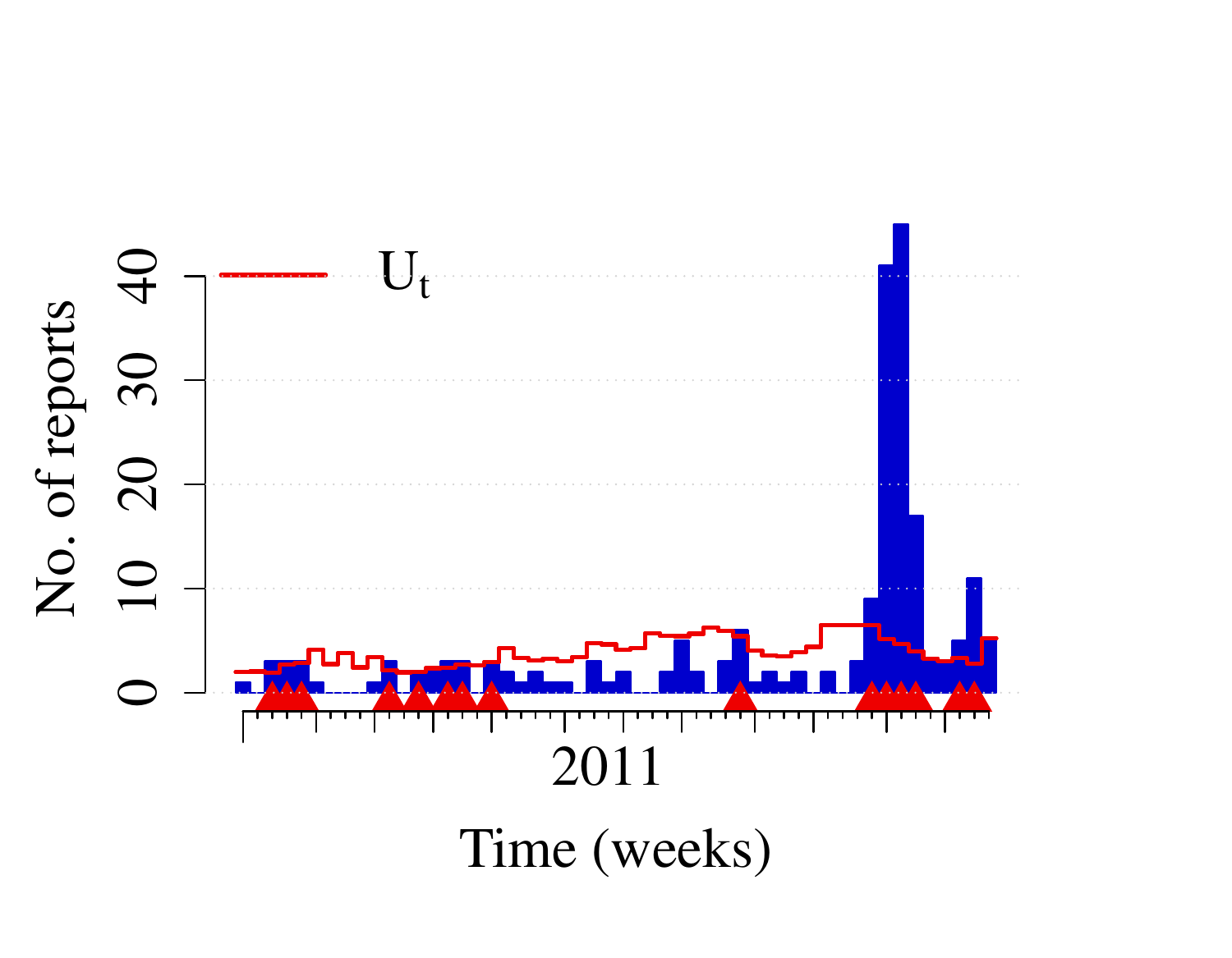}
}
\hspace{-3em}
\subfloat[]{
\includegraphics[width=9cm]{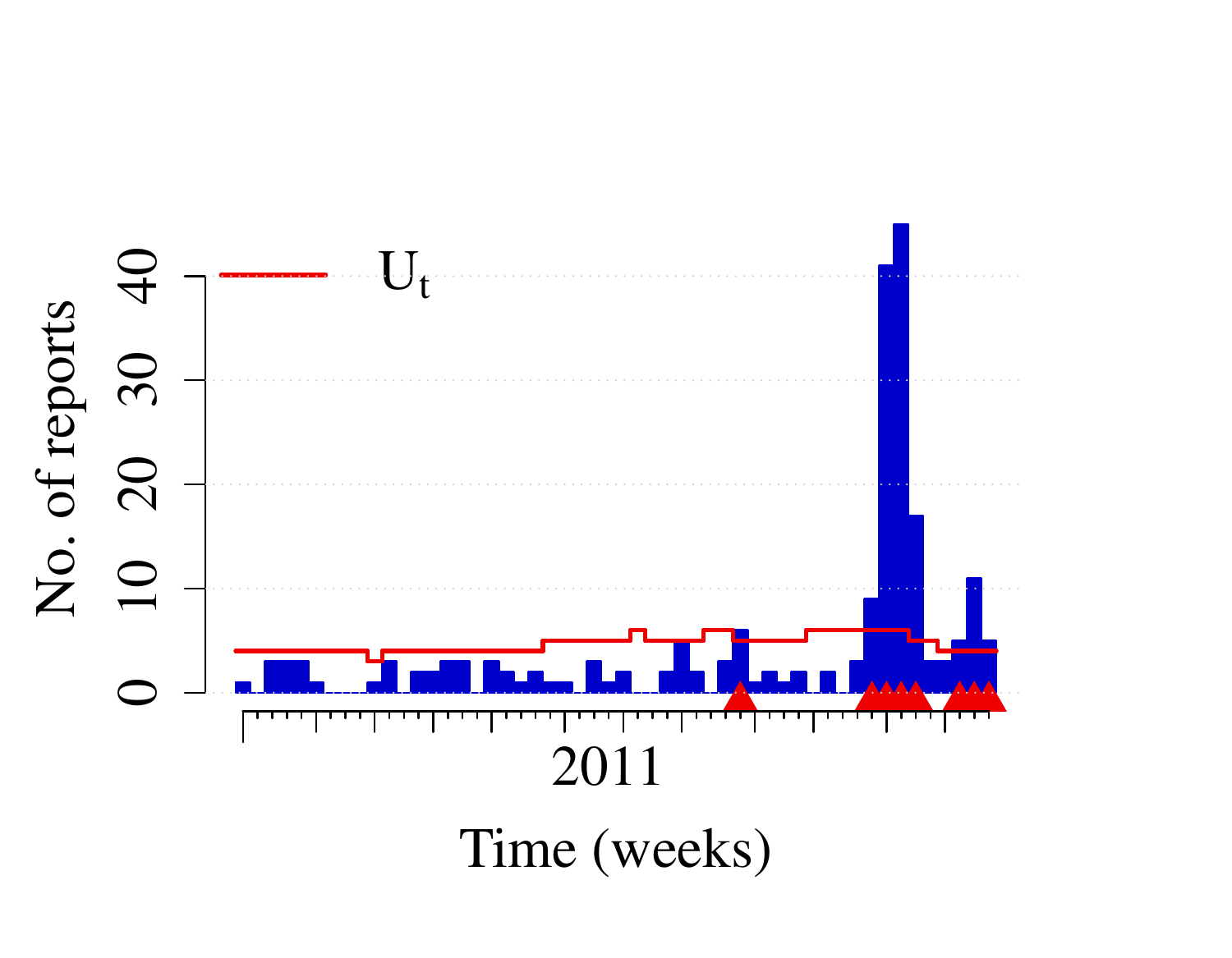}
}
\caption{S. Newport in Germany in 2011 monitored by (a) the original method and (b) the improved method. 
For the figure we turned off the option that the threshold is only computed if there were more than 5 cases during the 4 last timepoints including $t_0$.
One gets less alarms with the most recent method and still does not miss the outbreak in the summer. Simulations
 on more time series support the use of the improved method instead of the original method.}
 \label{fig:newportFar}
\end{figure}

The original method is widely used in public health surveillance~\citep{hulth_etal2010}. The reason for its success is primarily
 that it does not need to be fine-tuned for each specific pathogen. It is hence easy to implement it for scanning data for many different pathogens. Furthermore, it does tackle classical issues of surveillance data: overdispersion, presence of past outbreaks that are reweighted, 
seasonality that is taken into account differently in the two methods.
An example of use of the function is shown on Fig. \ref{fig:newportFar} with the code below.
\begin{Schunk}
\begin{Sinput}
R> salm.farrington <- farringtonFlexible(salmNewportGermany, control1)								  
R> salm.noufaily <- farringtonFlexible(salmNewportGermany, control2)
\end{Sinput}
\end{Schunk}

 \setkeys{Gin}{height=7cm, width=9cm}

With our implementation of the improvements presented in \citet{Noufaily2012} we hope that the method with time can replace the original method in routine use. The RKI system described in section~\ref{sec:RKI} already uses this improved method.
\subsubsection{Similar methods in the package}
The package also contains further methods based on a subset of the historical data: \code{bayes}, \code{rki} and \code{cdc}.
See Tab.~\ref{table:ref} for the corresponding references. Here, \code{bayes} uses a simple conjugate prior-posterior approach and computes the parameters of a 
  negative binomial distribution based on past values. The procedure \code{rki} makes either the assumption of a normal or a Poisson distribution based on the mean of past counts. Finally, \code{cdc} aggregates weekly data into 4-week-counts and computes a normal distribution based upper confidence interval. None of these methods offers the inclusion of a linear trend, down-weighting of past outbreaks or power transformation of the data.
 Although these methods are good to have at hand, we personally recommend the use of the improved method implemented in the function \code{farringtonFlexible}.
\subsection{A Bayesian refinement}
The \code{farringtonFlexible} function described previously 
was a first indication that the \textit{monitoring} of surveillance time
series requires a good \textit{modelling} of the time series before assessing
aberrations. Generalized linear models (GLMs) and generalized additive models
(GAMs) are well-established and powerful modelling frameworks for handling the
count data nature and trends of time series in a regression context. 
The \code{boda}  procedure~\citep{Manitz2013} continues this line of
thinking by extending the simple GLMs used in the \code{farrington} and
\code{farringtonFlexible} procedures to a fully fledged Bayesian GAM allowing
for penalized splines, e.g., to describe trends and seasonality, while 
simultaneously adjusting for previous outbreaks or concurrent
processes influencing the case counts.
A particular advantage of the Bayesian approach is that it constitutes a seamless
framework for performing both estimation and subsequent prediction: the 
uncertainty in parameter estimation is directly carried forward to the predictive
posterior distribution. No asymptotic normal approximations nor plug-in inference
is needed. For fast approximate Bayesian inference we use the \code{INLA} \proglang{R} 
package~\citep{INLA} to fit the Bayesian GAM. 

Still, monitoring with 
\code{boda} is substantially slower than using the Farrington procedures.
Furthermore, detailed regression modelling is only meaningful if the time series
is known to be subject to external influences on which information is available.
Hence, the typical use at a public health institution would be the 
detailed analysis of a few selected time series, \textit{e.g.}, critical ones or those
with known trend character. 

As an example,
\citet{Manitz2013} studied the influence of absolute
humidity on the occurence of weekly reported campylobacter cases in Germany.

\setkeys{Gin}{height=7cm, width=15cm}
\begin{figure}
\begin{center}  

\includegraphics{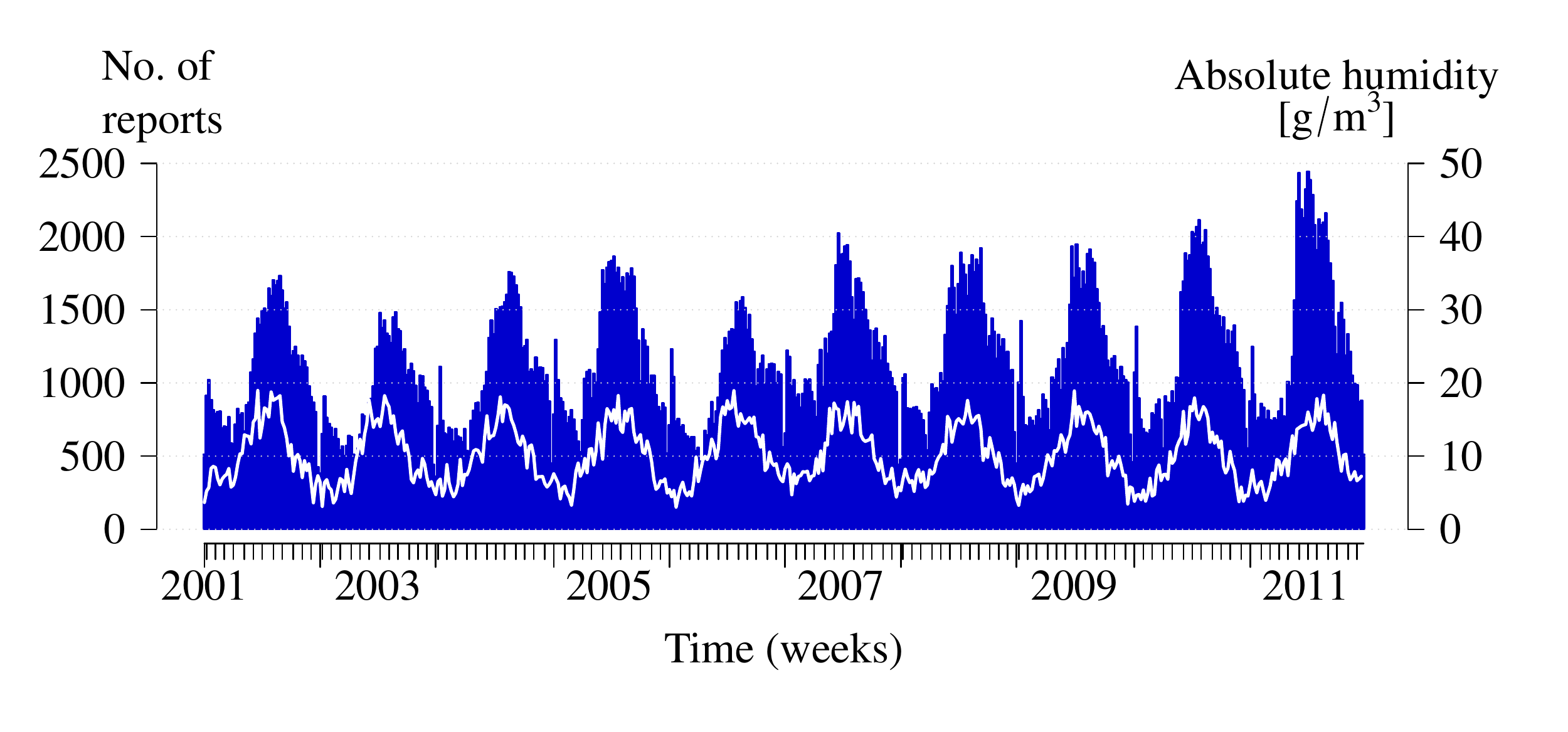}
\end{center}
\vspace{-1cm}
\caption{Weekly number of reported campylobacteriosis cases in Germany 2002-2011 as vertical bars. In addition, the corresponding mean absolute humidity time series is shown as a white curve.}
\label{fig:campyDE}
\end{figure}

\begin{Schunk}
\begin{Sinput}
R> data("campyDE")
R> cam.sts <- new("sts", epoch = as.numeric(campyDE$date),
                observed = campyDE$case, state = campyDE$state,
                epochAsDate = TRUE)		   
R> plot(cam.sts, legend = NULL, xlab = "time [weeks]", ylab = "No. reported",
      col = "gray", cex = 2, cex.axis = 2, cex.lab = 2)
R> lines(campyDE$hum * 50, col = "darkblue", lwd = 2)
\end{Sinput}
\end{Schunk}

The corresponding plot of the weekly time series is shown in
Fig.~\ref{fig:campyDE}. We observe a strong association between humidity 
and case numbers - an 
association which is stronger than with, \textit{e.g.}, temperature or relative humidity.
As noted in \citet{Manitz2013} the excess in cases in 2007 is thus partly explained by the high atmospheric humidity.
 Furthermore, an increase in case numbers during the 2011 STEC O104:H4 outbreak is observed, which is explained by increased awareness and testing of many gastroenteritits pathogens during that period. The hypothesis is thus that there is no actual increased disease 
activity~\citep{bernard_etal2014}. 
Unfortunately, the German reporting system only records positive test results without keeping track of the 
number of actual tests performed -- otherwise this would have been a natural adjustment variable. Altogether, the series contains several artefacts which 
appear prudent to address when monitoring the campylobacteriosis series.

The GAM in \code{boda} is based on the negative binomial distribution with time-varying expectation and time constant overdispersion parameter, \textit{i.e.}, 
\begin{align*}
y_t  &\sim \operatorname{NB}(\mu_t,\nu) 
\end{align*} 
with $\mu_{t}$ the mean of the distribution and $\nu$ the dispersion parameter~\citep{lawless1987}. 
To achieve compatibility with the quasi-Poisson model we let $\phi=\nicefrac{(\mu_{t}+\nu)}{\nu}$. Hence we have $\E(y_t)=\mu_t$ and $\Var(y_t)=\phi\cdot\mu_t$ with $\phi>1$. The linear predictor is given by
\begin{align*}
\log(\mu_t) &=  \alpha_{0t} + \beta t + \gamma_t + \bm{x}_t^\top \bm{\delta} + \xi z_t, \quad t=1,\ldots,t_0.
\end{align*}

Here, the time-varying intercept $\alpha_{0t}$ is described by a penalized spline
(\textit{e.g.}, first or second order random walk) and $\gamma_t$ denotes a periodic penalized
spline (as implemented in \code{INLA}) with period equal to the periodicity 
of the data. Furthermore, $\beta$ characterizes the effect of a possible linear
trend (on the log-scale) and $\xi$ is the effect of previous outbreaks. Typically, $z_t$ is
a zero-one process denoting if there was an outbreak in week $t$, but more involved adaptive and non-binary forms are imaginable. Finally, 
$\bm{x}_t$ denotes a vector of possibly time-varying covariates, which
influence the expected number of cases. Data from timepoints $1,\ldots,t_0-1$ are now used to determine the posterior distribution of all model parameters and subsequently the posterior predictive distribution of $y_{t_0}$ is computed. If the actual observed value of $y_{t_0}$ is above the $(1-\alpha)\cdot 100\%$ quantile of the predictive posterior distribution an alarm is flagged for $t_0$.

Below we illustrate the use 
of \code{boda} to monitor the campylobacterioris time series from 2007.
In the first case we include in the model for $\log\left(\mu_t\right)$ penalized splines for trend and 
seasonality and a simple linear trend.
\begin{Schunk}
\begin{Sinput}
R> rangeBoda <- which(epoch(cam.sts) >= as.Date("2007-01-01"))
R> control.boda <- list(range = rangeBoda, X = NULL, trend = TRUE,
                      season = TRUE, prior = "iid", alpha = 0.025, 
                      mc.munu = 10000, mc.y = 1000)
R> boda <- boda(cam.sts, control = control.boda)
\end{Sinput}
\end{Schunk}

In the second case we instead use only penalized and linear trend components, and,
furthermore, include as covariates lags 1-4 of the absolute humidity as well
as zero-one indicators for $t_0$ belonging to the last two weeks 
(\code{christmas}) or first two years (\code{newyears}) of the year,
respectively. The later two variables are needed, because there is a 
systematically changed reporting behaviour at the turn of the year (c.f.\ Fig.~\ref{fig:campyDE}). Finally, \code{O104period} is an indicator variable on
whether the reporting week belongs to the W21-W30 - 2011 period of increased awareness
during the O104:H4 STEC outbreak. No additional correction for past outbreaks is
made.
\begin{Schunk}
\begin{Sinput}
R> covarNames <- c("l1.hum", "l2.hum", "l3.hum", "l4.hum", 
                 "newyears", "christmas", "O104period")
R> control.boda2 <- modifyList(control.boda, 
                             list(X = campyDE[, covarNames], season = FALSE))
R> boda.covars <- boda(cam.sts, control = control.boda2)
\end{Sinput}
\end{Schunk}

\begin{figure}
\begin{center}  
  
\includegraphics{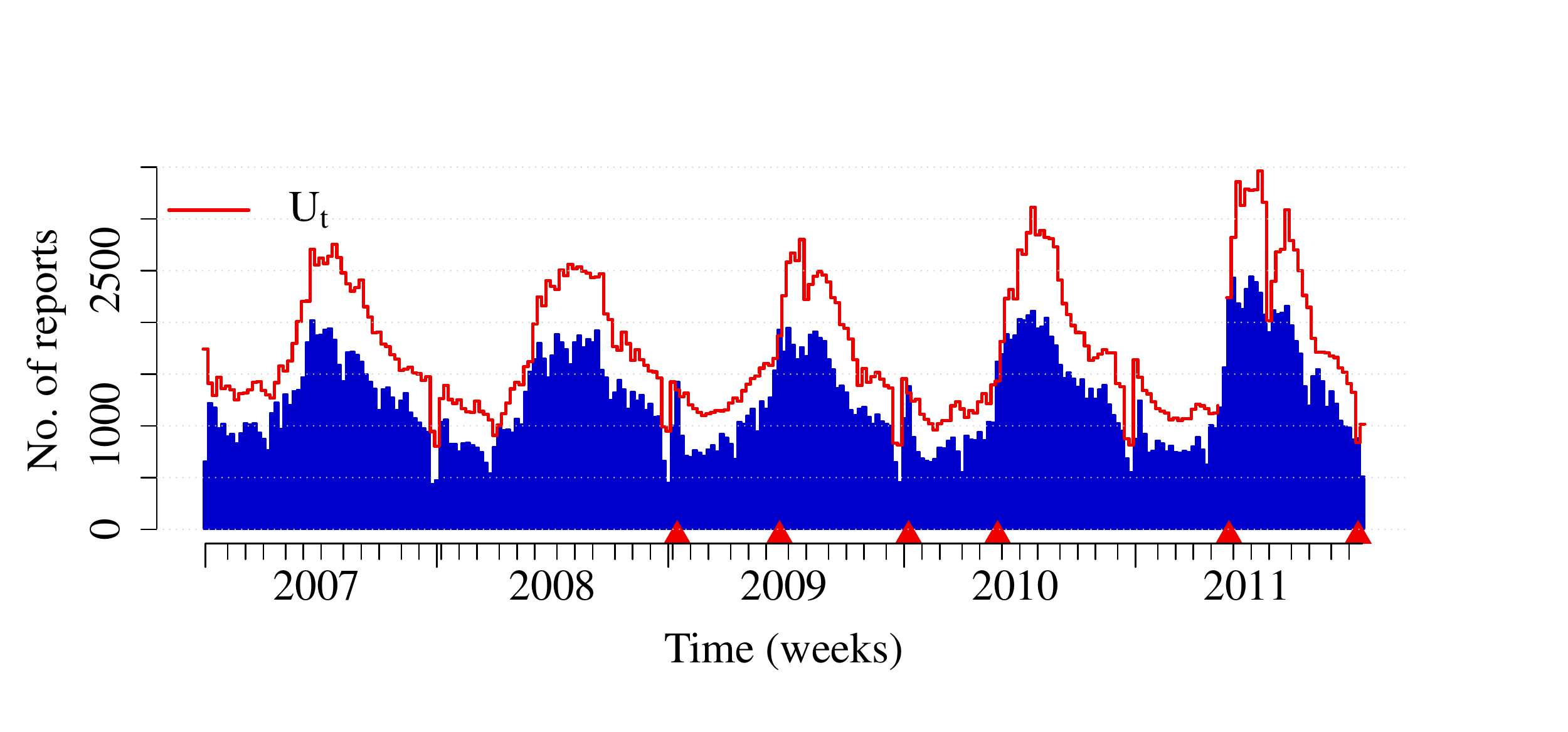}
\end{center}
\vspace{-1cm}
\caption{Weekly reports of Campylobacter in Germany in 2007-2011 monitored by the boda method with covariates. The line
represents the upperbound calculated by the algorithm. Triangles indicate alarms, \textit{i.e.} timepoints where the observed number of counts is higher than the upperbound.}
\label{fig:b}
\end{figure}

\begin{figure}
\begin{center}  
  
\includegraphics{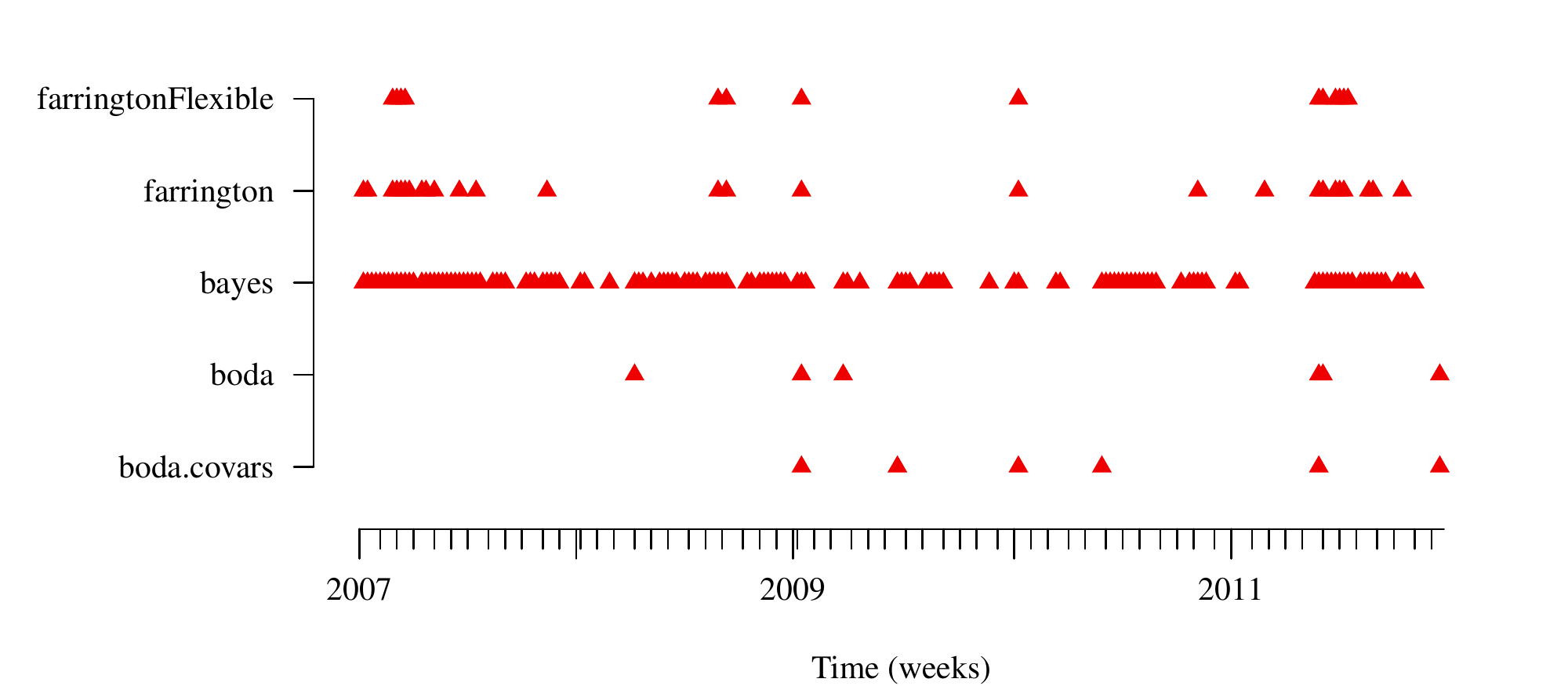}
\end{center}

\caption{Alarmplot showing the alarms for the campylobacteriosis time series for four different algorithms.}
\label{fig:alarmplot}
\end{figure}

We plot \code{boda.covars} in Fig.~\ref{fig:b} and compare the output of the two boda calls with the output of 
\code{farrington},  \code{farringtonFlexible} and \code{bayes} in 
Fig.~\ref{fig:alarmplot}. 

\begin{Schunk}
\begin{Sinput}
R> cam.surv <- combineSTS(list(boda.covars=boda.covars,boda=boda,bayes=bayes,
                             farrington=far,farringtonFlexible=farflex))                      
R> plot(cam.surv,type = alarm ~ time)
\end{Sinput}
\end{Schunk}

Note here that the Bayes
procedure is not really useful as the adjustment for seasonality only works
poorly. Moreover, we think that this method produces many false alarms for this time series because it disregards the increasing time trend in number of 
reported cases. Furthermore, it becomes clear that the improved Farrington procedure acts
similar to the original procedure, but the improved reweighting and trend
inclusion produces fewer alarms.

 \setkeys{Gin}{height=7cm, width=15cm}

\setkeys{Gin}{height=7cm, width=16cm}
\subsection{Beyond one-timepoint detection}
GLMs as used in the Farrington method are suitable for the purpose of aberration detection since they allow a regression approach for adjusting counts for known phenomena such as trend or seasonality in surveillance data. 
Nevertheless, the Farrington method only performs one-timepoint detection. In some contexts it can be more relevant 
to detect sustained shifts early, \textit{e.g.}, an outbreak could be characterized at first by counts slightly higher than usual 
in subsequent weeks without each weekly count being flagged by one-timepoint detection methods. Control charts inspired by statistical process control (SPC) 
\textit{e.g.}, cumulative sums would allow the detection of sustained shifts. Yet they were not tailored the specific characteristics of surveillance data such as overdispersion or seasonality. 
The method presented in \citet{hoehle_paul2008} conducts a synthesis of both worlds, \textit{i.e.} traditional surveillance methods and SPC. The method is implemented in the package as the function
\code{glrnb}, whose use is explained here.
\subsubsection{Definition of the control chart}
For the control chart, two distributions are defined, one for each of the two states \textit{in-control} and \textit{out-of-control}, whose likelihoods are compared at each time step. The \textit{in-control} distribution
$f_{\bm{\theta}_0}(y_t|\bm{z}_t)$ with the covariates $\bm{z}_t$ is estimated by a GLM of the Poisson or negative binomial family with a log link, depending on the overdispersion of the data. 
In this context, the standard model for the \textit{in-control} mean is
$$\log \mu_{0,t}=\beta_0+\beta_1t+\sum_{s=1}^S\left[\beta_{2s}\cos \left(\frac{2\pi s t}{\texttt{Period}}\right)+\beta_{2s+1}\sin \left(\frac{2\pi s t}{\texttt{Period}}\right)\right] $$ 
where $S$ is the number of harmonic waves to use and \texttt{Period} is the period of the data as indicated in the \code{control} slot, for instance 52 for weekly data. 
But more flexible linear predictors, \textit{e.g.}, containing splines, concurrent covariates 
or an offset could be used on the right hand-side of the equation. 
The GLM could therefore be made very close
 to the one used by~\citet{Noufaily2012}, with reweighting of past outbreaks and various criteria for including the time trend. 
 
 The parameters of the \textit{in-control} and \textit{out-of-control} models 
 are respectively given by $\bm{\theta}_0$ and $\bm{\theta}_1$.
 The \textit{out-of-control} mean is defined as a function of the \textit{in-control} mean, either with a multiplicative shift (additive on the log-scale) whose size $\kappa$ can be given as an input 
 or reestimated at each timepoint $t>1$, $\mu_{1,t}=\mu_{0,t}\cdot \exp(\kappa)$, or with an unknown autoregressive component as in \citet{held_etal2005}, $\mu_{1,t}=\mu_{0,t}+\lambda y_{t-1}$
 with unknown $\lambda>0$.
 
 In \code{glrnb}, timepoints are divided into two intervals: phase 1 and phase 2. The \textit{in-control} mean and overdispersion are estimated with a GLM fitted on phase 1 data, whereas surveillance operates on phase 2 data. 
 When $\lambda$ is fixed, one uses a likelihood-ratio and
 defines the stopping time for alarm as
$$N=\min \left\{ t_0 \geq 1 : \max_{1\leq t \leq t_0} \left[ \sum_{s=t}^{t_0} \log\left\{ \frac{f_{\bm{\theta}_1}(y_s|\bm{z}_s)}{f_{\bm{\theta}_0}(y_s|\bm{z}_s)} \right\} \right] \geq \texttt{c.ARL} \right\},$$
where $n$ is the number of timepoints in the phase 2 data and $\texttt{c.ARL}$ is the threshold of the CUSUM.

When $\lambda$ is unknown and with the autoregressive component one has to use a generalized likelihood ration (GLR) with the following stopping rule to estimate them on the fly at each time point so that
$$N_G=\min \left\{ t_0 \geq 1 : \max_{1\leq t \leq t_0} \sup_{\bm{\theta} \in \bm{\Theta}} \left[ \sum_{s=t}^{t_0} \log\left\{ \frac{f_{\bm{\theta}}(y_s|\bm{z}_s)}{f_{\bm{\theta}_0}(y_s|\bm{z}_s)} \right\} \right] \geq \texttt{c.ARL} \right\}\cdot$$
Thus, one does not make any hypothesis about the specific value of the change to detect, but this GLR is more computationally intensive than the LR.
\subsubsection{Practical use}
For using \code{glrnb} one has two choices to make. First, one has to choose an \textit{in-control} model that will be fitted on phase 1 data. One can either provide the
 predictions for the vector of \textit{in-control} means \code{mu0} and the overdispersion parameter \code{alpha} by relying on an external fit, or use the built-in GLM estimator,
  that will use all data before the beginning of the surveillance range to fit a GLM with the number of harmonics \code{S} and a time trend if \code{trend} is \code{TRUE}.
The choice of the exact \textit{in-control} model depends on the data under surveillance. Performing model selection is a compulsory step in practical applications.

Then, one needs to tune the surveillance function itself, for one of the two possible change forms, \code{intercept} or \code{epi}. 
One can choose either to set \code{theta} to a given value and thus perform LR instead of GLR. The value of \code{theta} has to be adapted to 
the specific context in which the algorithm is applied: how big are shifts one wants to detect optimally? Is it better not to specify any and use GLR instead? 

The threshold \texttt{c.ARL} also has to be specified by the user. As explained in \citet{hoehle_mazick2009} one can compute the threshold for
 a desired run-length in control through direct Monte Carlo simulation or a Markov chain approximation. Lastly, as mentioned in
  \citet{hoehle_paul2008}, a window-limited approach of surveillance, instead of looking at all the timepoints until the first observation, can make computation faster. 

\begin{figure}
\begin{center}  
  
\includegraphics{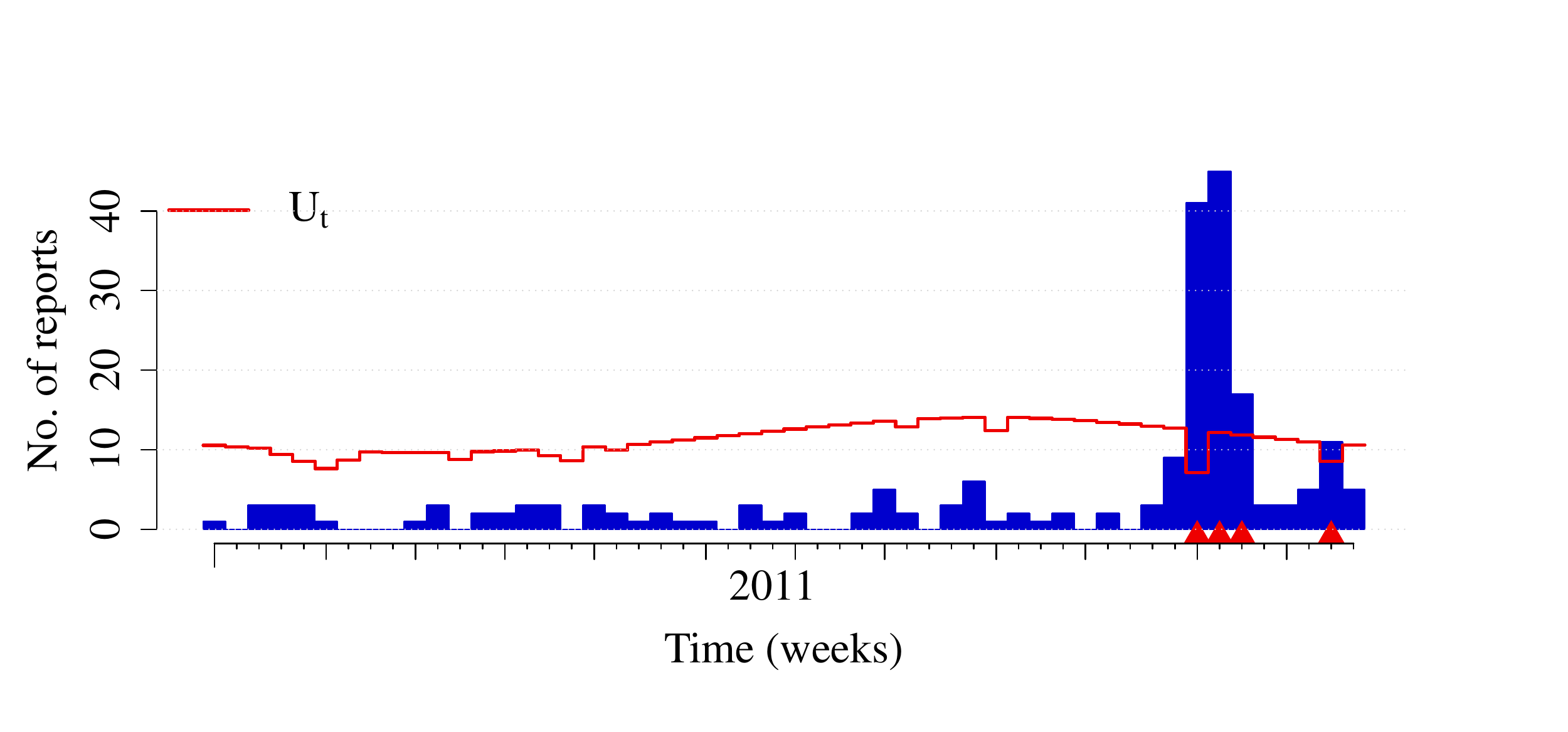}
\end{center}
\vspace{-1cm}
\caption{S. Newport in Germany in 2011 monitored by the \code{glrnb} function. }
\label{fig:glrnb}
\end{figure}

Here we apply \code{glrnb} to the time series of report counts of \textit{Salmonella Newport} in Germany by assuming a known multiplicative shift of factor $2$ and by using the built-in estimator to fit an \textit{in-control} model with one harmonic for 
   seasonality and a trend. This model will be refitted after each alarm, but first we use data from the years before 2011 as reference or \code{phase1}, 
   and the data from 2011 as data to be monitored or \code{phase2}. The threshold \texttt{c.ARL} was chosen to be 4 as we found with the same approach as \citet{hoehle_mazick2009} that it made the probability of a false alarm within one year 
   smaller than 0.1. Figure ~\ref{fig:glrnb} shows the results of this monitoring.

\begin{Schunk}
\begin{Sinput}
R> phase1 <- which(isoWeekYear(epoch(salmNewportGermany))$ISOYear < 2011)
R> phase2 <- in2011
R> control = list(range = phase2, c.ARL = 4, theta = log(2), ret = "cases",
                mu0 = list(S = 1, trend = TRUE, refit = FALSE))		 
R> salmGlrnb <- glrnb(salmNewportGermany, control = control)
\end{Sinput}
\end{Schunk}
 \setkeys{Gin}{height=7cm, width=15cm}

The implementation of \code{glrnb} on individual time series was already thoroughly explained in \citet{hoehle_mazick2009}. Our objective in the present document is rather to provide practical tips 
for the implementation of this function on huge amounts of data in public health surveillance applications. Issues of computational speed become very significant in such a context. Our proposal 
to reduce the computational burden incurred by this algorithm is 
to compute the \textit{in-control} model for each time serie (pathogen, subtype, subtype in a given location, \textit{etc.}) only once a year and to use this estimation for the computation of a threshold for each time series. 
 An idea to avoid starting with a 0 CUSUM is to use either $\left(\nicefrac{1}{2}\right)\cdot\texttt{c.ARL}$ as a starting value (fast initial response 
 CUSUM as presented in~\citet{lucas1982fast}) or to let surveillance run with the new \textit{in-control} model during 
 a buffer period and use the resulting CUSUM as an initial value. One could also choose the maximum of these two possible starting values as a starting value.
 During the buffer period alarms would be generated with the old model. Lastly, using GLR is much more computationally intensive than using LR, 
 whereas LR performs reasonably well on shifts different from the one indicated by \code{theta} as seen in the simulation studies of~\citet{hoehle_paul2008}. Our advice would therefore be to use LR with a reasonable predefined \code{theta}. 
 The amount of historical data used each year to update the model, the length of the buffer period and the value of \code{theta} have to be fixed for each specific application, \textit{e.g.} using simulations and/or 
 discussion with experts.

\subsubsection{Similar methods in the package}
The algorithm \code{glrPois} is the same function as \code{glrnb} but for Poisson distributed data.
Other CUSUM methods for count data are found in the package: \code{cusum} and \code{rogerson}. Both methods are discussed and compared to \code{glrnb} in  \citet{hoehle_paul2008}.
The package also includes a semi-parametric method \code{outbreakP} that aims at detecting changes from a constant level to a monotonically increasing incidence, for instance the beginning of the influenza season. See Tab.~\ref{table:ref} for the corresponding references.
\subsection{A method for monitoring categorical data}
All monitoring methods presented up to now have been methods for analysing count data. Nevertheless, in public health surveillance one also encounters categorical time series 
which are time series where the response variable obtains one of $k\geq2$ different categories (nominal or ordinal). When $k=2$ the time series is binary, for instance representing 
a specific outcome in cases such as hospitalization, death or a positive result to some diagnostic test. One can also think of applications with 
$k>2$ if one studies \textit{e.g.}, the age groups of the cases in the context of monitoring a vaccination program: vaccination targeted at children could induce a shift towards older cases which one wants to detect 
as quickly as possible -- this will be explained thorougly with an example.

The developments of prospective surveillance methods for such categorical time series were up to recently limited to CUSUM-based approaches for 
binary data such as those explained in~\citet{Chen1978},~\citet{Reynolds2000} and~\citet{rogerson_yamada2004}. Other than being only suitable for binary data these methods have the drawback of not handling 
overdispersion. A method improving on these two limitations while casting the problem into a more comprehending GLM regression framework for categorical data was 
presented in~\citet{hoehle2010}. It is implemented as the function
 \code{categoricalCUSUM}.
 
The way \code{categoricalCUSUM} operates is very similar to what \code{glrnb} does with fixed \textit{out-of-control} parameter.
First, the parameters in a multivariate GLM for the \textit{in-control} distribution are estimated from the historical data. Then the \textit{out-of-control} distribution is defined by a given 
change in the parameters of this GLM, \textit{e.g.}, an intercept change, as explained later. Lastly, prospective monitoring is performed on 
current data using a likelihood ratio detector which compares the likelihood of the response under the \textit{in-control} and \textit{out-of-control} distributions. 
\subsubsection{Categorical CUSUM for Binomial Models}
The challenge when performing these steps with categorical data from surveillance systems 
is finding an appropriate model. Binary GLMs as presented in Chapter 6 of~\citet{fahrmeir2013regression} could be a solution but they do not tackle well the inherent overdispersion in the binomial time series. 
Of course one could choose a quasi family but these are not proper statistical distributions making many issues such as prediction complicated. A better alternative 
is offered by the use of \textit{generalized additive models for location, scale and shape} (GAMLSS)~\citep{Rigby2005},  that support distributions such as the beta-binomial distribution, suitable for overdispersed binary data. With GAMLSS one can model the dependency of the mean -- \textit{location}  -- 
upon explanatory variables but the regression modelling is also extended to other parameters of the distribution, \textit{e.g.} scale. Moreover any modelled parameter can be put under surveillance, be it the mean (as in the example later developed) 
or the time trend in the linear predictor of the mean. This very flexible modelling framework is implemented in \proglang{R} through the \pkg{gamlss} package~\citep{StasJSS}.

\begin{figure}
\begin{center}  
  
\includegraphics{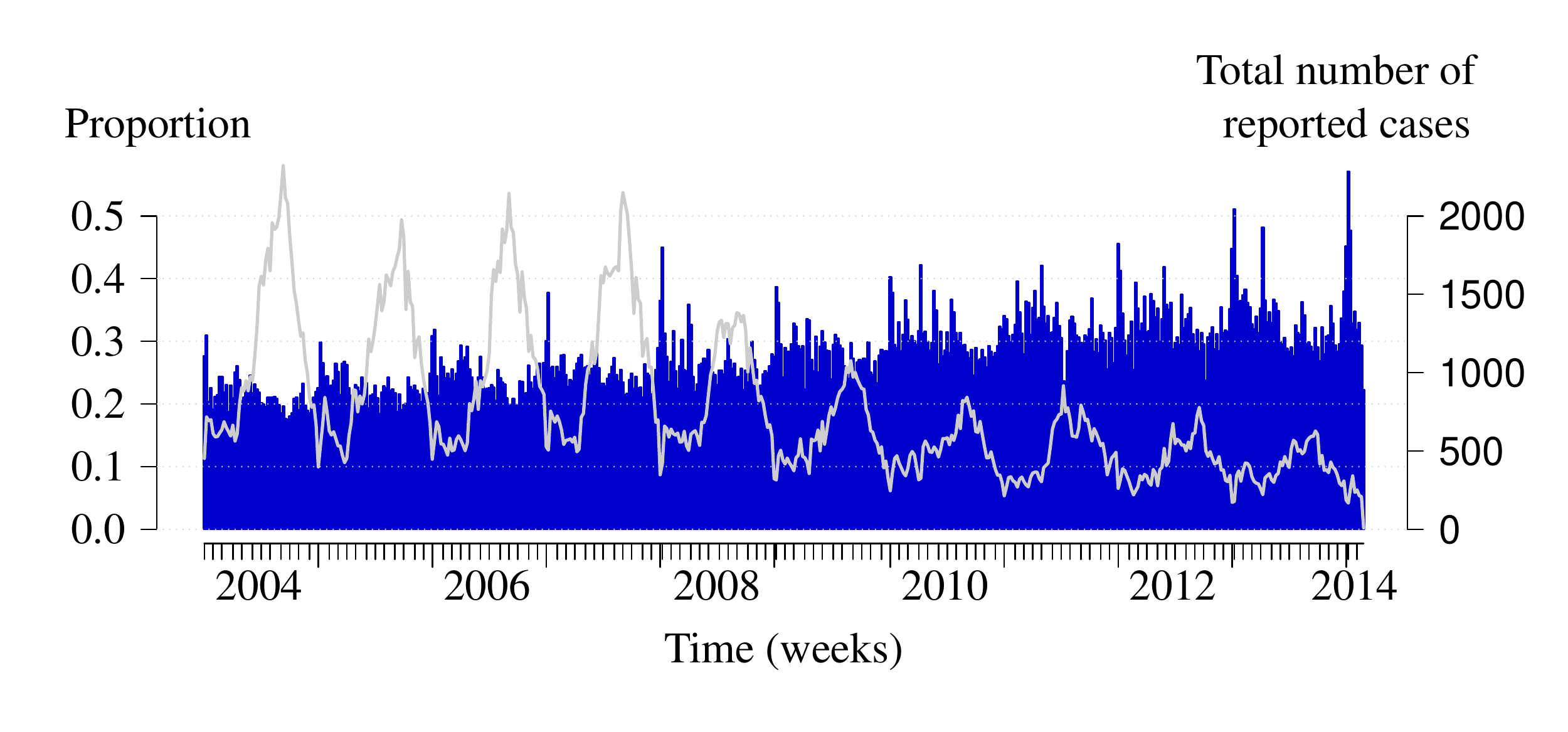}
\end{center}
\vspace{-1cm}
\caption{Weekly proportion of Salmonella cases that were hospitalized in Germany 2004-2014. In addition the corresponding number of reported cases is shown as a light curve.}
\label{fig:cat1}
\end{figure}

As an example we consider the time series of the weekly number of hospitalized cases among all \textit{Salmonella} cases in Germany in Jan 2004-Jan 2014, depicted on Fig.~\ref{fig:cat1}. We use 2004-2012 data to estimate the \textit{in-control} parameters and then perform surveillance on the data from 2013 and early 2014. We start by preprocessing the data. 
\begin{Schunk}
\begin{Sinput}
R> data("salmHospitalized")
R> isoWeekYearData <- isoWeekYear(epoch(salmHospitalized))
R> dataBefore2013 <- which(isoWeekYearData$ISOYear < 2013)
R> data2013 <- which(isoWeekYearData$ISOYear == 2013)
R> dataEarly2014 <- which(isoWeekYearData$ISOYear == 2014
                       & isoWeekYearData$ISOWeek <= 4)
R> phase1 <- dataBefore2013
R> phase2 <- c(data2013, dataEarly2014)
R> weekNumbers <- isoWeekYearData$ISOWeek
R> salmHospitalized.df <- cbind(as.data.frame(salmHospitalized), weekNumbers)
R> colnames(salmHospitalized.df) <- c("y", "t", "state", "alarm", "n",
                                    "freq", "epochInPeriod", "weekNumber")
\end{Sinput}
\end{Schunk}

We assume that the number of hospitalized cases follows a beta-binomial distribution, \textit{i.e.},
$ y_t \sim \BetaBin(n_t,\pi_t,\sigma_t)$ with $n_t$ the total number of reported cases at time $t$, $\pi_t$ the proportion of these cases that were hospitalized and $\sigma$ the dispersion parameter. In this 
parametrization, $$E(y_t)=n_t \pi_t,\quad \text{and}$$ $$\Var(y_t)=n_t \pi_t(1-\pi_t)\left( 1 + \frac{\sigma(n_t-1)}{\sigma+1} \right)\cdot$$
We choose to model the expectation $n_t \pi_t$ using a beta-binomial model with a logit-link which is a special case of a GAMLSS, \textit{i.e.},
$$\logit(\pi_t)=\bm{z}_t^\top\bm{\beta}$$
where $\bm{z}_t$ is a vector of possibly time-varying covariates and $\bm{\beta}$ a vector of covariate effects in our example.

The proportion of hospitalized cases 
varies throughout the year as seen in Fig.~\ref{fig:cat1}.  
One observes that in the summer the proportion of hospitalized cases is smaller than in other seasons. However, over the holidays in December the proportion of hospitalized cases increases.  
 Note that the number of non-hospitalized cases drops while the number of hospitalized cases remains constant (data not shown): this might be explained by the fact that cases that are not serious enough to go to the hospital are not seen by general practitioners because sick workers do not need 
 a sick note during the holidays. Therefore, the \textit{in-control} model should contain these elements, as well as the fact that there is an 
increasing trend of the proportion because GPs prescribe less and less stool diagnoses so that more diagnoses are done on hospitalized cases.

We choose a model with an intercept, a time trend, two harmonic terms and a factor variable for the first two weeks of 
each year. The variable \code{epochInPeriod} takes into account the fact that not all years have 52 weeks. 
\begin{Schunk}
\begin{Sinput}
R> vars <- c( "y", "n", "t", "epochInPeriod", "weekNumber")
R> m.bbin <- gamlss(cbind(y, n-y) ~ 1 + t
                  + sin(2 * pi * epochInPeriod) + cos(2 * pi * epochInPeriod) 
                  + sin(4 * pi * epochInPeriod) + cos(4 * pi * epochInPeriod)
                  + I(weekNumber == 1) + I(weekNumber == 2), 
                  sigma.formula =~ 1,
                  family = BB(sigma.link = "log"),
                  data = salmHospitalized.df[phase1, vars])
\end{Sinput}
\end{Schunk}

The change we aim to detect is defined by a multiplicative change of odds, from $\nicefrac{\pi_t^0}{(1-\pi_t^0)}$ to $R\cdot\nicefrac{\pi_t^0}{(1-\pi_t^0)}$ with $R>0$, similar to what was done in~\citet{Steiner1999} for the logistic regression model. 
This is equivalent to an additive change of the log-odds,
$$\logit(\pi_t^1)=\logit(\pi_t^0)+\log R$$
with $\pi_t^0$ being the \textit{in-control} proportion and $\pi_t^1$ the \textit{out-of-control} distribution. The likelihood ratio based CUSUM statistic is now defined as
$$C_{t_0}=\max_{1\leq t \leq {t_0}}\left( \sum_{s=t}^{t_0} \log \left( \frac{f(y_s;\bm{z}_s,\bm{\theta}_1)}{f(y_s;\bm{z}_s,\bm{\theta}_0)} \right) \right)$$
with $\bm{\theta}_0$ and $\bm{\theta}_1$ being the vector in- and \textit{out-of-control} parameters, respectively. Given a threshold \code{h}, an alarm is sounded at the first time when $C_{t_0}>\texttt{h}$. 

We set the parameters of the \code{categoricalCUSUM} to optimally detect a doubling of the odds in 2013 and 2014, \textit{i.e.} $R=2$. Furthermore, we for now set the threshold of the CUSUM at $h=2$. 
We use the GAMLSS to predict the mean of the \textit{in-control} and \textit{out-of-control} distributions and store them into matrices with two columns among which the second one represents the reference category.
\begin{Schunk}
\begin{Sinput}
R> R <- 2 
R> h <- 2
R> pi0 <- predict(m.bbin, newdata = salmHospitalized.df[phase2, vars],
                type = "response")
R> pi1 <- plogis(qlogis(pi0) + log(R))
R> pi0m <- rbind(pi0, 1 - pi0)
R> pi1m <- rbind(pi1, 1 - pi1)
\end{Sinput}
\end{Schunk}
Note that the \code{categoricalCUSUM} function is constructed to operate on the observed slot of \code{sts}-objects 
which have as columns the number of cases in each category at each timepoint, \textit{i.e.}, each row of the observed slot contains the elements 
$(y_{t1},...,y_{tk})$. 

\begin{Schunk}
\begin{Sinput}
R> populationHosp <- cbind(population(salmHospitalized), 
                         population(salmHospitalized))
R> observedHosp <- cbind(observed(salmHospitalized), 
                       population(salmHospitalized) - 
 					  observed(salmHospitalized))
R> nrowHosp <- nrow(salmHospitalized)
R> salmHospitalized.multi <- new("sts", freq = 52, start = c(2004, 1),
                               epoch = as.numeric(epoch(salmHospitalized)),
                               epochAsDate = TRUE,
                               observed = observedHosp,
                               populationFrac = populationHosp,
                               state = matrix(0, nrow = nrowHosp, ncol = 2), 
                               multinomialTS = TRUE)
\end{Sinput}
\end{Schunk}
Furthermore, one needs to define a wrapper for the distribution function in order to have a argument named \code{"mu"} in the function. 
\begin{Schunk}
\begin{Sinput}
R> dBB.cusum <- function(y, mu, sigma, size, log = FALSE) {
 	 return(dBB(if (is.matrix(y)) y[1, ] else y, 
 				if (is.matrix(y)) mu[1, ] else mu,
 				sigma = sigma, bd = size, log = log)) 
 }
\end{Sinput}
\end{Schunk}

\begin{figure}
\hspace{-1em}
\subfloat[]{

\includegraphics[width=9cm]{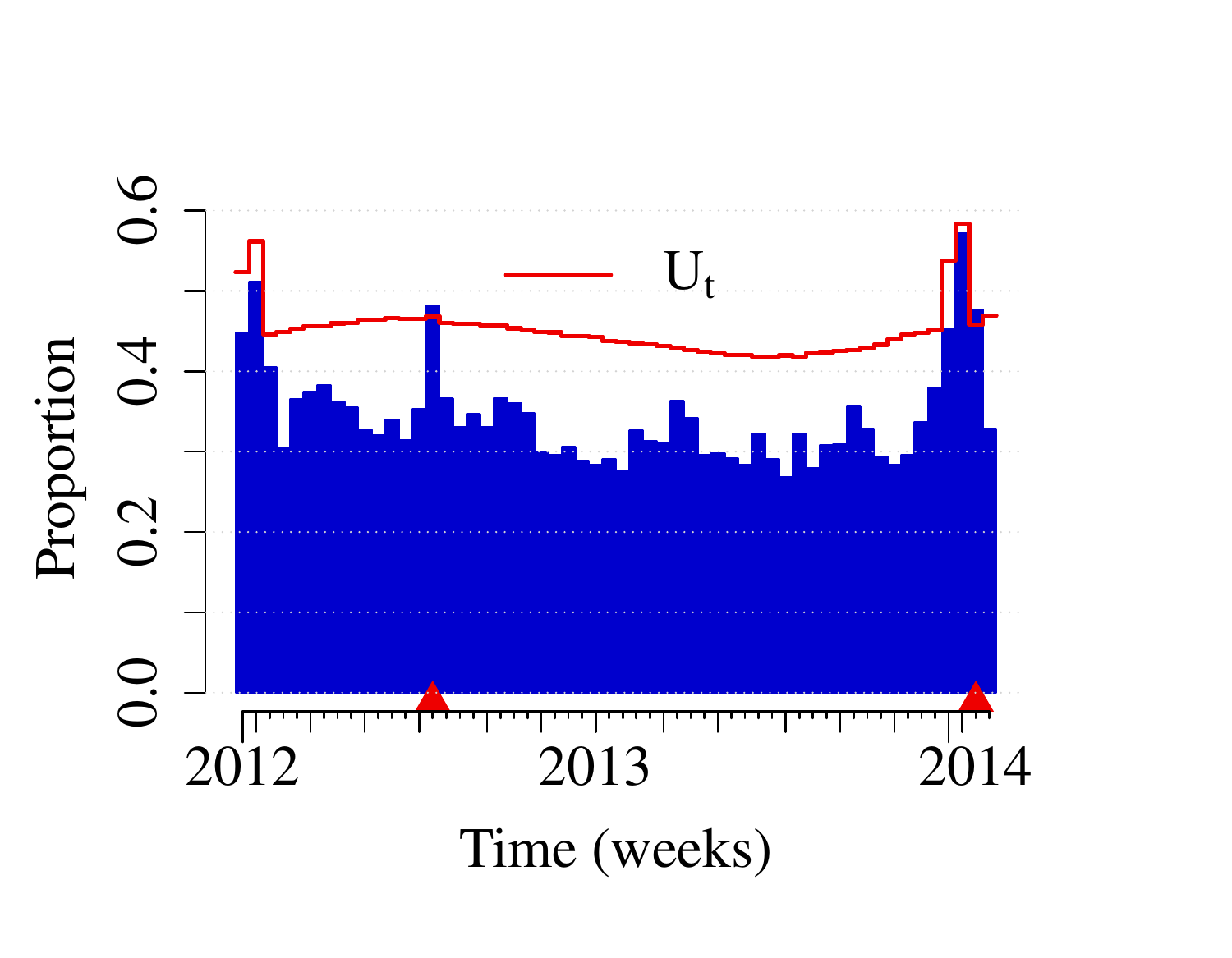}
}
\hspace{-3em}
\subfloat[]{
\includegraphics[width=9cm]{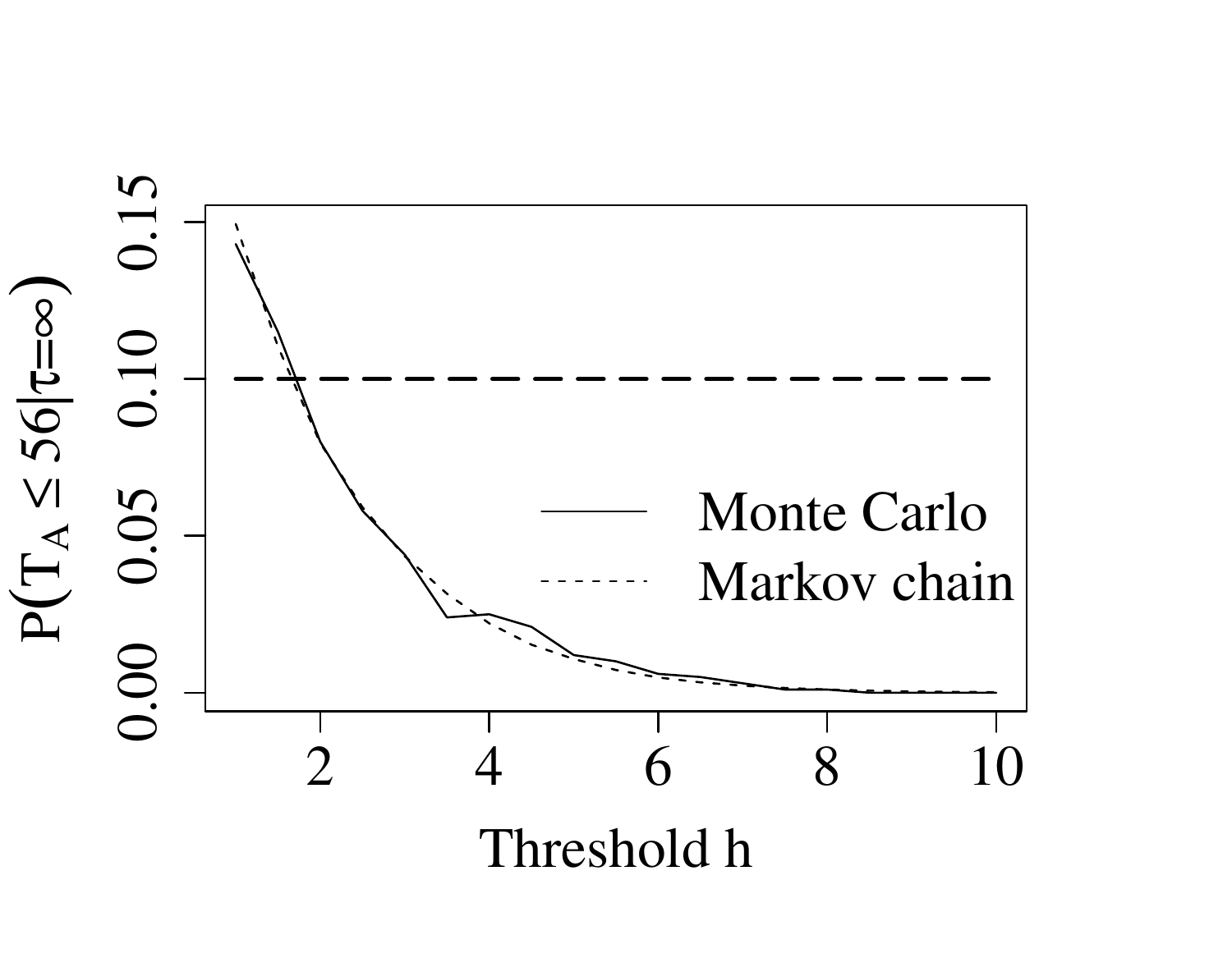}
}
\caption{(a) Results of the monitoring with categoricalCUSUM of the proportion of Salmonella cases that were hospitalized in Germany in Jan 2013 - Jan 2014. (b) 
Probability of a false alarm within the 56 timepoints of the monitoring as a function of the threshold $h$.}
\label{fig:catDouble}

\end{figure}

After these preliminary steps, the monitoring can be performed.
\begin{Schunk}
\begin{Sinput}
R> controlCat <- list(range = phase2, h = 2, pi0 = pi0m, pi1 = pi1m, 
                 ret = "cases", dfun = dBB.cusum)
R> salmHospitalizedCat <- categoricalCUSUM(salmHospitalized.multi, 
                                         control = controlCat,
                                         sigma = exp(m.bbin$sigma.coef))									
\end{Sinput}
\end{Schunk}
The results can be seen in Fig.~\ref{fig:catDouble}(a). With the given settings, there are alarms at week 
16 in 2004 
and at week 3 in 2004. 
The one in 2014 corresponds to the usual peak of the beginning of the year, which was larger than expected this year, maybe because the 
weekdays of the holidays were particularly worker-friendly so that sick notes were even less needed.

The value for the threshold \code{h} can be determined following the procedures presented in \citet{hoehle_mazick2009} for count data, and as in the code exhibited below. Two methods can be used for 
determining the probability of a false alarm within a pre-specified number of steps for a given value of the threshold \code{h}: a Monte Carlo method relying on \textit{e.g.}, 1000 simulations and a Markov Chain approximation of the CUSUM. The former is much more computationally intensive than the latter: 
with the code below, the Monte Carlo method needed approximately 300 times more time than the Markov Chain method.
Since both results are close we recommend the Markov Chain approximation for practical use. The Monte Carlo method works by sampling observed values from the estimated 
distribution and performing monitoring with \code{categoricalCUSUM} on this \code{sts}-object. As observed values are estimated from the \textit{in-control} distribution every alarm thus obtained is a 
false alarm so that the simulations allow to estimate the probability of a false alarm when monitoring \textit{in-control} data over the timepoints of \code{phase2}. The Markov Chain approximation introduced by \citet{brook_evans1972} is implemented as \code{LRCUSUM.runlength} 
which is already used for \code{glrnb}. Results from both methods can be seen on figure 
\ref{fig:catDouble}(b). We chose a value of 2 for \code{h} so that the probability of a false alarm within the 56 timepoints of \code{phase2} is less than $0.1$.

One first has to set the values of the threshold to be investigated and to prepare the function used for simulation, that draws observed values from the 
\textit{in-control} distribution and performs monitoring on the corresponding time series, then indicating if there was at least one alarms. Then 1000 simulations were 
performed with a fixed seed value for the sake of reproducibility. Afterwards, we tested the Markov Chain approximation using the function \code{LRCUSUM.runlength} over the same grid 
of values for the threshold.
\begin{Schunk}
\begin{Sinput}
R> h.grid <- seq(1, 10, by = 0.5)			
R> simone <- function(sts, h) {
   y <- rBB(length(phase2), mu = pi0m[1, , drop = FALSE],
 			 bd = population(sts)[phase2, ],
 	         sigma = exp(m.bbin$sigma.coef))
   observed(sts)[phase2, ] <- cbind(y, sts@populationFrac[phase2, 1] - y) 
   one.surv <- categoricalCUSUM(sts, modifyList(controlCat, list(h = h)),
 	                           sigma = exp(m.bbin$sigma.coef))
   return(any(alarms(one.surv)[, 1]))
 }
R> set.seed(123)
R> nSims <- 1000
R> pMC <- sapply(h.grid, function(h) { 
 	mean(replicate(nSims, simone(salmHospitalized.multi, h))) 
 })
R> pMarkovChain <- sapply( h.grid, function(h) {
   TA <- LRCUSUM.runlength(mu = pi0m[1,, drop = FALSE], 
                           mu0 = pi0m[1,, drop = FALSE], 
                           mu1 = pi1m[1,, drop = FALSE],
                           n = population(salmHospitalized.multi)[phase2, ],
                           h = h, dfun = dBB.cusum,
                           sigma = exp(m.bbin$sigma.coef))
   return(tail(TA$cdf, n = 1))
 })
\end{Sinput}
\end{Schunk}

\setkeys{Gin}{height=7cm, width=9cm}

The procedure for using the function for multicategorical variable follows the same steps (as illustrated later). Moreover, one could expand the approach 
to utilize the multiple regression possibilities offered by GAMLSS. Here we chose to try to detect a change in the mean of the distribution of counts but as GAMLSS provides more general regression tools 
than GLM we could also aim at detecting a change in the time trend included in the model for the mean. 

\subsubsection{Categorical CUSUM for Multinomial Models}

In order to illustrate the use of \code{categoricalCUSUM} for more than two classes we analyse the monthly number of rotavirus cases in the federal state Brandenburg during 2002-2013 and which are stratified into the five age-groups 00-04, 05-09, 10-14, 15-69, 70+ years. In 2006 two rotavirus vaccines were introduced, which
are administered in children at the age of 4-6 months. Since then,
coverage of these vaccination has steadily increased and interest is to detect possible age-shifts in the distribution of cases.
\begin{Schunk}
\begin{Sinput}
R> data("rotaBB")
R> plot(rotaBB, xlab = "Time (months)", 
      ylab = "Proportion of reported cases")
\end{Sinput}
\end{Schunk}

 \setkeys{Gin}{height=7cm, width=15cm}
\begin{figure}
\includegraphics{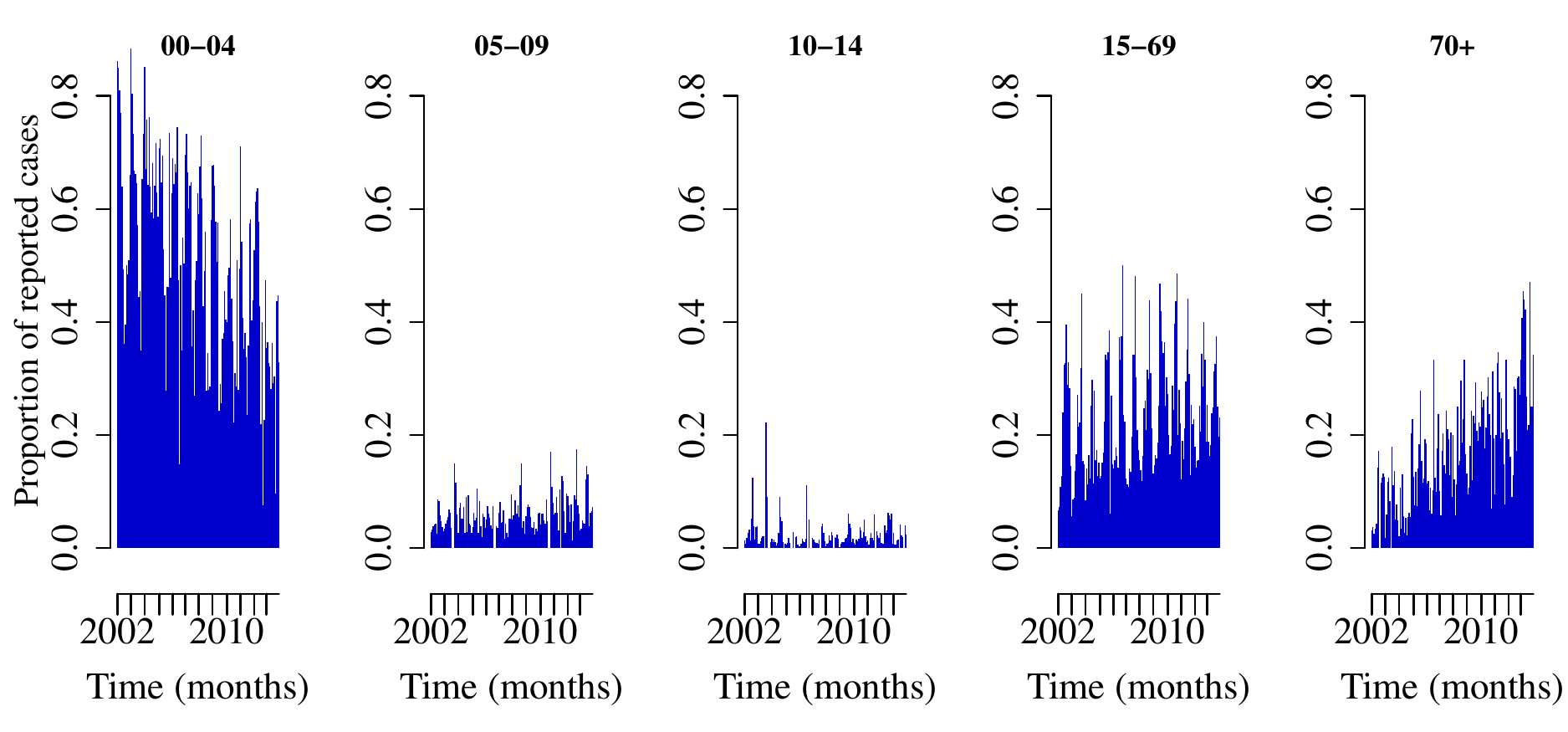}
\caption{Monthly proportions in five age-groups for the reported rotavirus cases in Brandenburg, Germany, 2002-2013.}
\label{fig:vac}
\end{figure}
From Fig.~\ref{fig:vac} we observe a shift in proportion away from the very young. However, interpreting the proportions only makes sense in combination with the absolute numbers. In these plots (not shown) it becomes clear that the absolute numbers in the 0-4 year old have decreased since 2009. However, in the 70+ group a small increase is observed with 2013 by far being the strongest season so far.
Hence, our interest is in prospectively detecting a possible age-shift. Since the vaccine was recommended for routine vaccination in Brandenburg in 2009 we choose to start the monitoring at that time point. We do so by fitting a multinomial logit-model containing a trend as well as one harmonic wave and use the age group 0-4 years as reference category, 
to the data from the years 2002-2008. Different \proglang{R} packages implement such type of modelling, but we shall use the \pkg{MGLM} package~\citet{MGLM}, because it also offers the fitting of extended multinomial regression models allowing for extra dispersion.

\begin{Schunk}
\begin{Sinput}
R> rotaBB.df <- as.data.frame(rotaBB)
R> X <- with(rotaBB.df, cbind(intercept = 1, epoch,
                            sin1 = sin(2 * pi * epochInPeriod), 
                            cos1 = cos(2 * pi * epochInPeriod)))
R> phase1 <- epoch(rotaBB) < as.Date("2009-01-01") 
R> phase2 <- !phase1
R> order <- c(2:5, 1); reorder <- c(5, 1:4)
R> library("MGLM")
R> m0 <- MGLMreg(as.matrix(rotaBB.df[phase1, order]) ~ -1 + X[phase1, ], 
               dist = "MN")               
\end{Sinput}
\end{Schunk}

As described in \citet{hoehle2010} we can try to detect a specific shift in the intercept coefficients of the model. For example, a multiplicative shift of factor 7 in the example below, in the odds of each of the four age categories against the reference category is modelled by changing the intercept value of each category. 
Based on this, the \textit{in-control} and \textit{out-of-control} proportions are easily computed using the \code{predict} function for \code{MGLMreg} objects.

\begin{Schunk}
\begin{Sinput}
R> m1 <- m0
R> m1$coefficients[1, ] <- m0$coefficients[1, ] + log(7)
R> pi0 <- t(predict(m0, newdata = X[phase2, ])[, reorder])
R> pi1 <- t(predict(m1, newdata = X[phase2,])[, reorder])
\end{Sinput}
\end{Schunk}

For applying the \code{categoricalCUSUM} function one needs to define a compatible wrapper function for the multinomial as in the binomial example.

With $\bm{\pi}^0$ and $\bm{\pi}^1$ in place one only needs to define a wrapper function, which defines the PMF of the sampling distribution -- in this case the multinomial -- in a \code{categoricalCUSUM} compatible way. 
\begin{Schunk}
\begin{Sinput}
R> dfun <- function(y, size, mu, log = FALSE) {
 	return(dmultinom(x = y, size = size, prob = mu, log = log))
 }
R> control <- list(range = seq(nrow(rotaBB))[phase2], h = h, pi0 = pi0, 
 				pi1 = pi1, ret = "value", dfun = dfun)
R> surv <- categoricalCUSUM(rotaBB,control=control)
\end{Sinput}
\end{Schunk}

\begin{figure}
\hspace{-1em}
\subfloat[]{

\includegraphics[width=9cm]{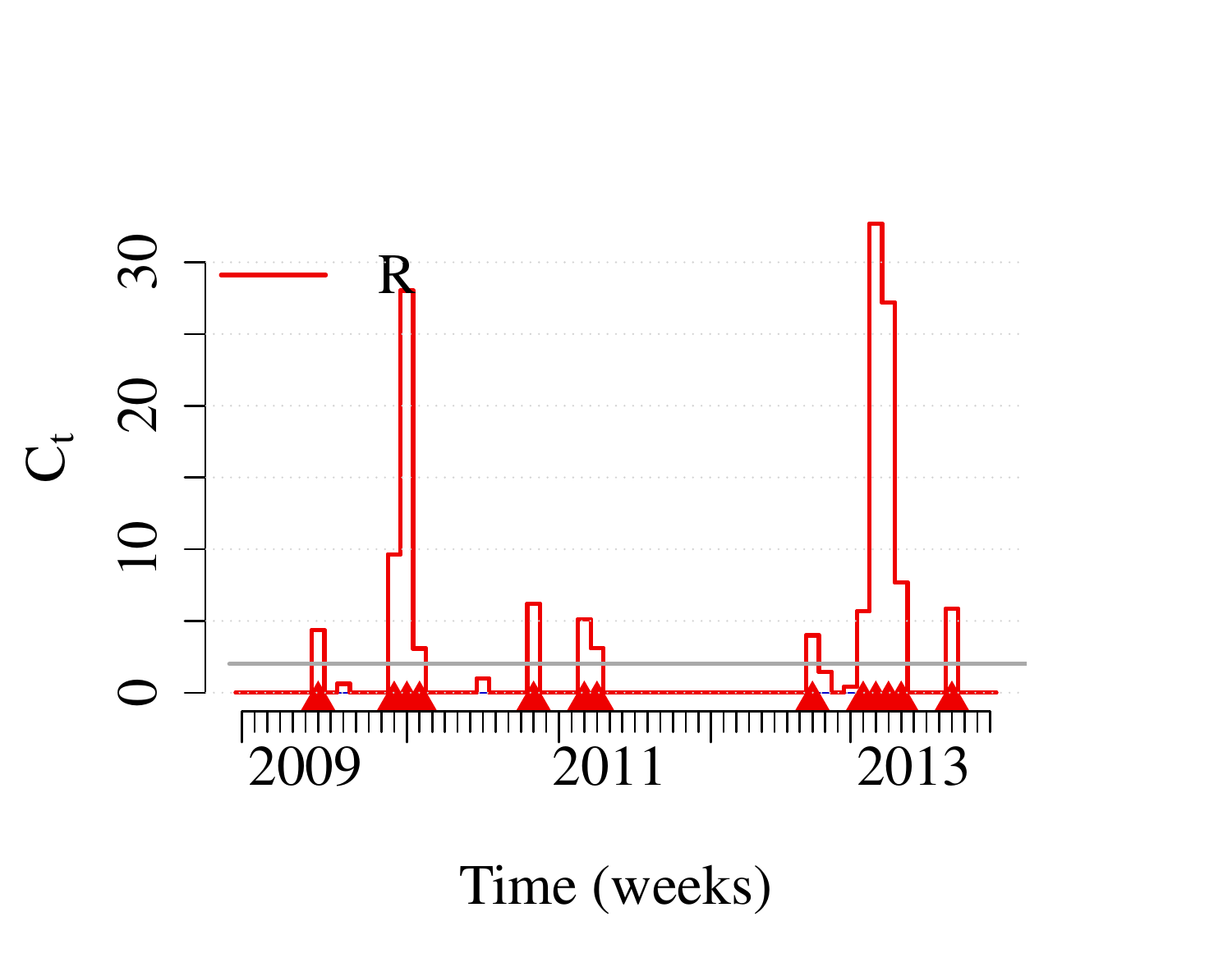}
}
\hspace{-3em}
\subfloat[]{
\includegraphics[width=9cm]{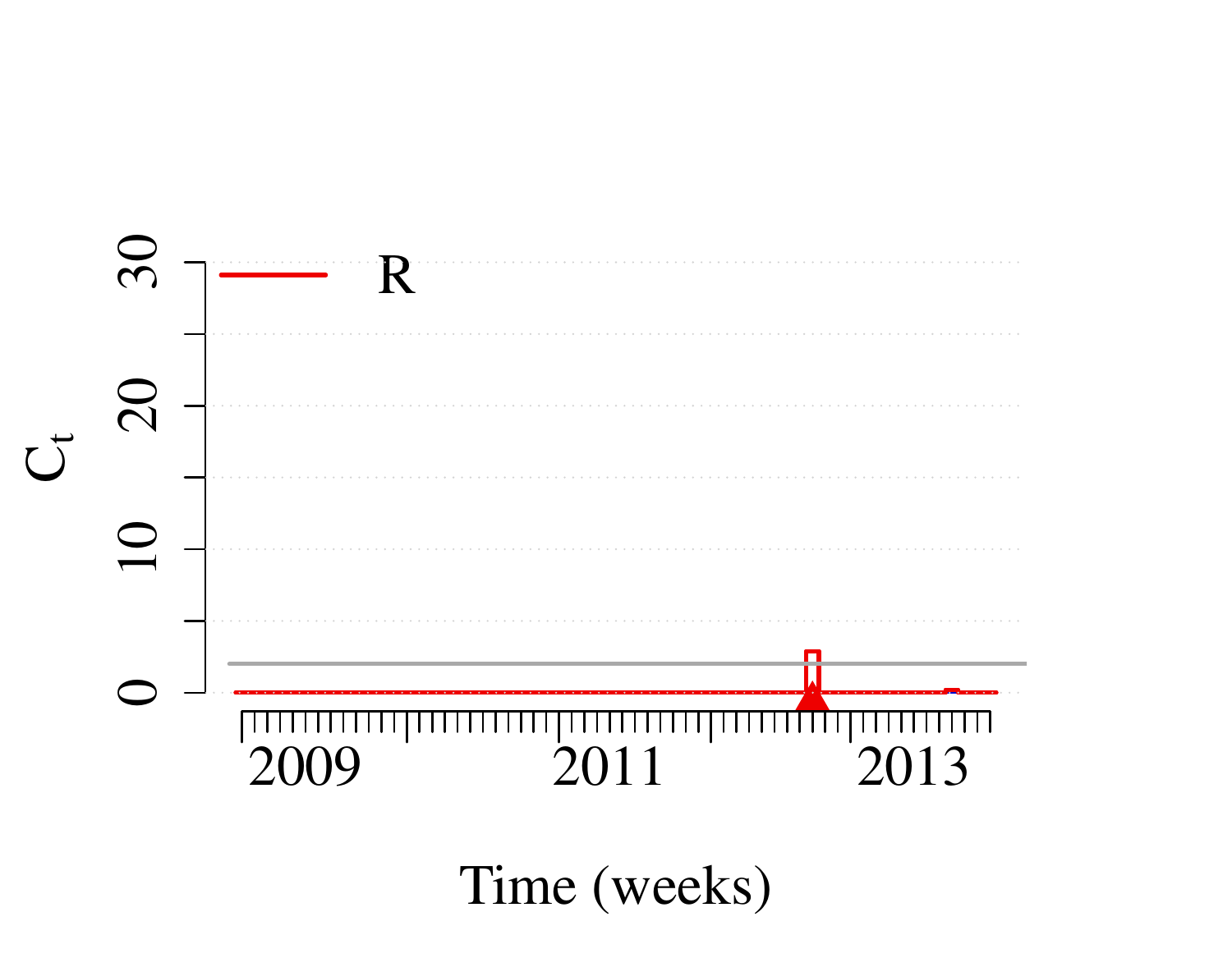}
}
\caption{Categorical CUSUM statistic $C_t$. Once $C_t>2$ an alarm is sounded and the statistic is reset. In (a) surveillance uses the multinomial distribution and in (b) surveillance uses the Dirichlet-multinomial distribution.}
\label{fig:ct}

\end{figure}

The resulting CUSUM statistic $C_t$ as a function of time is shown in Fig.~\ref{fig:ct}(a). The first time an aberration is detected is July 2009. Using 10000 Monte Carlo simulations we estimate that with the chosen threshold $h=2$ the probability for a false alarm within the 60 time points of \code{phase2} is 0.02.

As the above example shows, the LR based categorical CUSUM is rather flexible in handling any type of multivariate GLM modelling to specify the \textit{in-control} and \textit{out-of-control} proportions. However, it requires a direction of the change to be specified -- for which detection is optimal. One sensitive part of such monitoring is the fit of the multinomial distribution to a multivariate time series of proportions, which usually exhibit extra dispersion when compared to the multinomial. For example comparing the AIC between the multinomial logit-model and a Dirichlet-multinomial model with $\alpha_{ti} = \exp(\bm{x}_t^\top\bm{\beta})$~\citep{MGLM} shows that overdispersion is present. 
The Dirichlet distribution is the multicategorical equivalent of the beta-binomial distribution. We exemplify its use in the code below. 

\begin{Schunk}
\begin{Sinput}
R> m0.dm <- MGLMreg(as.matrix(rotaBB.df[phase1, 1:5]) ~ -1 + X[phase1, ], 
                 dist = "DM")
R> c(m0$AIC, m0.dm$AIC)
\end{Sinput}
\begin{Soutput}
[1] 2485.446 2060.532
\end{Soutput}
\end{Schunk}

Hence, the above estimated false alarm probability might be too low for the actual monitoring problem, because the variation in the time series is larger than implied by the multinomial. Hence, it appears prudent to repeat the analysis using the more flexible  Dirichlet-multinomial model. This is straightforward with \code{categoricalCUSUM} once the \textit{out-of-control} proportions are specified in terms of the model. Such a specification is, however, hampered by the fact that the two models use different parametrizations.

For performing monitoring in this new setting we first need to calculate the $\alpha$'s of the multinomial-Dirichlet for the \textit{in-control} and \textit{out-of-control} 
distributions.

\begin{Schunk}
\begin{Sinput}
R> delta <- 2
R> m1.dm <- m0.dm 
R> m1.dm$coefficients[1, ] <- m0.dm$coefficients[1, ] + 
                            c(-delta, rep(delta/4, 4))
R> alpha0 <- exp(X[phase2,] %*% m0.dm$coefficients)
R> alpha1 <- exp(X[phase2,] %*% m1.dm$coefficients)
R> dfun <- function(y, size, mu, log = FALSE) {
 	dLog <- ddirm(t(y), t(mu))
 	if (log) { return(dLog) } else { return(exp(dLog)) } 
 }
R> h <- 2
R> control <- list(range = seq(nrow(rotaBB))[phase2], h = h, 
                 pi0 = t(alpha0), pi1 = t(alpha1), 
                 ret = "value", dfun = dfun)
R> surv.dm <- categoricalCUSUM(rotaBB, control = control)
\end{Sinput}
\end{Schunk}

\setkeys{Gin}{height=7cm, width=9cm}

The resulting CUSUM statistic $C_t$ using the Dirichlet multinomial distribution is shown in Fig.~\ref{fig:ct}(b). 
We notice a rather similar behaviour even though the shift-type specified by this model is slightly different than in the model of Fig.~\ref{fig:ct}(a).   
\subsubsection{Categorical data in routine surveillance}

The multidimensionality of data available in public health surveillance creates many opportunities for the application of categorical time series: one could \textit{e.g.}, look at the sex ratio of cases of a given disease, at the age categories distribution, 
at the regions sending data, \textit{etc.} If one is interested in monitoring with respect to a categorical variable, a choice has to be made between monitoring each time series individually, 
for instance a time series of \textit{Salmonella} cases for each age category, or to monitor the distribution of cases with respect to that factor \textit{via} \code{categoricalCUSUM}. A downside of the latter 
solution is that one has to specify the change parameter \code{R} in advance. However, more straightforward applications could be found in the surveillance of positive diagnostics if one were to obtain data about tests performed by laboratories and not only about confirmed cases. An alternative would be to apply 
\code{farringtonFlexible} while using the number of tests as \code{populationOffset}.

\subsubsection{Similar methods in the package}
The package also offers another CUSUM method suitable for binary data, \code{pairedbinCUSUM} that implements the method introduced by~\citet{Steiner1999}, which does not, however, take overdispersion into account as well as \code{glrnb}. The algorithm \code{rogerson} also supports the analysis of binomial data. See Tab.~\ref{table:ref} for the corresponding references.
\bigskip
\subsection{Other algorithms implemented in the package}
We conclude this description of surveillance methods by giving an overview of all algorithms implemented in the package with the corresponding references in Tab.~\ref{table:ref}. 
One can refer to the relative reference articles and to the reference
 manual of the package for more information about each method.
 
 Criteria
for choosing a method in practice are numerous. First one needs to ponder on the amount of historical data at hand -- for instance the EARS methods only need data for the last
timepoints whereas the Farrington methods use data up to $b$ years in the past. Then one should consider the amount of past data used by the algorithm -- historical reference methods use only a subset of the past data, namely 
the timepoints located around the same timepoint in the past years, whereas other methods use all past data included in the reference data.
This can be a criterion of choice since one can prefer using all available data. It is also important to decide whether one wants to detect one-timepoint aberration or more prolonged shifts.
And lastly, an important criteria is how much work needs to be done for finetuning the algorithm for each specific time series. 

The package on the one hand provides the means for analysing nearly all type of surveillance data 
and on the other hand makes the comparison of algorithms possible. This is useful in practical applications when those algorithms are implemented into routine use, which
will be the topic of section~\ref{sec:routine}.
\begin{table}[!h]

\centering

\begin{tabularx}{15cm}{|X|X|}
\hline
\pkg{surveillance} function & References \\
\hline
\code{boda} & \citet{Manitz2013}  \\
\hline
\code{bayes}   & \citet{riebler2004} \\
\hline
\code{categoricalCUSUM} & \citet{hoehle2010}\\
\hline
\code{cdc}   & \citet{stroup_etal89,farrington2003} \\
\hline
\code{cusum}  & \citet{rossi_etal99,pierce_schafer86} \\
\hline
\code{earsC} & \citet{SIM:SIM3197} \\
\hline
\code{farrington} & \citet{farrington96} \\
\hline
\code{farringtonFlexible} & \citet{farrington96,Noufaily2012} \\
\hline
\code{glrnb} & \citet{hoehle_paul2008} \\
\hline
\code{glrpois} & \citet{hoehle_paul2008} \\
\hline
\code{outbreakP} & \citet{frisen_etal2009,fri2009} \\
\hline
\code{pairedbinCUSUM} & \citet{Steiner1999} \\
\hline
\code{rki}  & Not available -- unpublished   \\
\hline
\code{rogerson} & \citet{rogerson_yamada2004}   \\
\hline
\end{tabularx}

\caption{Algorithms for aberration detection implemented in \pkg{surveillance}.}
\label{table:ref}
\end{table}

%% file: part3.tex
\DefineVerbatimEnvironment{Sinput}{Verbatim}{fontshape=sl}
\DefineVerbatimEnvironment{SinputSmall}{Verbatim}{fontshape=sl}
\DefineVerbatimEnvironment{Soutput}{Verbatim}{fontshape=s1}
\DefineVerbatimEnvironment{Scode}{Verbatim}{fontshape=sl}
\setkeys{Gin}{width=0.9\textwidth}
\indent  Combining \pkg{surveillance} with other \proglang{R} packages and programs is easy, allowing the integration of
the aberration detection into a comprehensive surveillance system to be used in routine practice. In our opinion, such a surveillance system has to at least support the following process: loading data from local databases, analysing them within \pkg{surveillance} 
and sending the results of this analysis to the end-user who is typically an epidemiologist in charge of the specific pathogen. This section exemplifies the integration of the package 
into a whole analysis stack, first through the introduction of a simple workflow from data query to a \code{Sweave} report of signals, and
 secondly through the presentation of the more elaborate system in use at the German Robert Koch Institute.\\
\subsection{A simple surveillance system}
Suppose you have a database with surveillance time series but little resources to build a surveillance system encompassing all the above stages. Using \proglang{R} and \code{Sweave} for \LaTeX~you can still set up
 a simple surveillance analysis without having to do everything by hand. You only need to input the data into \proglang{R} and create \code{sts}-objects for each time series of interest 
 as explained thoroughly in~\citet{hoehle_mazick2009}. Then, after choosing a surveillance algorithm, say \code{farringtonFlexible}, and 
 feeding it with the appropriate \code{control} argument, you can get a sts-object with upperbounds and alarms for each of your time series of interest over the \code{range} 
 supplied in \code{control}. For defining the range automatically one could use the R-function \code{SysDate()} to get today's date. 
 These steps can be introduced as a code chunk in a \code{Sweave} code that will translate it into a report that you can send to the epidemiologists in charge of the respective pathogen whose cases are monitored. 
 
 Below is an example of a short code segment showing the analysis of the \textit{S. Newport} weekly counts of cases in the German federal states Baden-W\"{u}rttemberg and North Rhine-Westphalia 
with the improved method implemented in \code{farringtonFlexible}. The package provides a \code{toLatex} method for \code{sts} objects that produces a table with the observed number of counts and upperbound for each column in 
\code{observed}, where alarms can be highlighted by for instance bold text. The resulting table is shown in Tab.~\ref{tableResults}.
\begin{Schunk}
\begin{Sinput}
R> data("salmNewport")
R> today <- which(epoch(salmNewport) == as.Date("2013-12-23"))
R> rangeAnalysis <- (today - 4):today
R> in2013 <- which(isoWeekYear(epoch(salmNewport))$ISOYear == 2013)
R> algoParameters <- list(range = rangeAnalysis, noPeriods = 10, 
                        populationBool = FALSE,
                        b = 4, w = 3, weightsThreshold = 2.58,
                        pastWeeksNotIncluded = 26, pThresholdTrend = 1,
                        thresholdMethod = "nbPlugin", alpha = 0.05,
                        limit54 = c(0, 50))
R> results <- farringtonFlexible(salmNewport[, c("Baden.Wuerttemberg",
                                              "North.Rhine.Westphalia")],
                   control = algoParameters)
R> start <- isoWeekYear(epoch(salmNewport)[range(range)[1]])
R> end <- isoWeekYear(epoch(salmNewport)[range(range)[2]])
R> caption <- paste("Results of the analysis of reported S. Newport 
                  counts in two German federal states for the weeks W-",
                  start$ISOWeek, "-", start$ISOYear, " - W-", end$ISOWeek,
                  "-", end$ISOYear, " performed on ", Sys.Date(),
                  ". Bold upperbounds (UB) indicate weeks with alarms.", 
                  sep="")
R> toLatex(results, caption = caption)
\end{Sinput}
\end{Schunk}
\begin{table}[h]
\centering
{\normalsize
\begin{tabular}{rrrlrr}
  \hline
Year & Week & Baden-Wuerttemberg & Threshold & North-Rhine-Westphalen & Threshold \\ 
  \hline
2013 & 48 & 0 & 5 & 0 & 4 \\ 
  2013 & 49 & 1 & 3 & 0 & 3 \\ 
  2013 & 50 & 1 & 3 & 0 & 3 \\ 
  2013 & 51 & 2 & 3 & 0 & 3 \\ 
  2013 & 52 & 3 & \textbf{\textcolor{red}{2}} & 1 & 3 \\ 
   \hline
\end{tabular}
}
\caption{Results of the analysis of reported S. Newport 
                 counts in two German federal states for the weeks W-48 2013 - W-52 2013 performed on 2014-07-29. Bold upperbounds (thresholds) indicate weeks with alarms.} 
\label{tableResults}
\end{table}
The advantage of this approach is that it can be made automatic. The downside of such a system is that the report is not interactive, for instance one cannot click on the cases and get the linelist. Nevertheless, this is a workable solution in 
many cases -- especially when human and financial resources are narrow. 
In the next Section, we present a more advanced surveillance system built on the package.

\subsection{Automatic detection of outbreaks at the Robert Koch Institute}
\label{sec:RKI}
The package \pkg{surveillance} was used as a core building block for designing and implementing the automated outbreak detection system at the RKI in Germany~\citep{Dirk}.
Due to the Infection Protection Act (IfSG) the RKI daily receives over 1,000 notifiable disease reports. The system analyses about half a million time series per day to identify possible aberrations in the reported number of cases. 
Structurally, it consists of two components: an analytical process written in \proglang{R} that daily monitors the data and a reporting component that compiles and communicates the results to the epidemiologists.

The analysis task relies on \pkg{surveillance} and three other \proglang{R} packages, namely \pkg{data.table}, \pkg{RODBC} and \pkg{testthat} as described in the following. The data-backend is an OLAP-system~\citep{SSAS} and relational databases, which are queried using \pkg{RODBC}~\citep{rodbc2013}.
 The case reports are then fastly aggregated into univariate time series using \pkg{data.table}~\citep{datatable2013}. To each time series we apply the \code{farringtonFlexible} algorithm on univariate \code{sts}-objects and store the analysis results in another SQL-database.
We make intensive use of \pkg{testthat}~\citep{testthat2013} for automatic testing of the component.
Although \proglang{R} is not the typical language to write bigger software components for production, choosing \proglang{R} in combination with \pkg{surveillance} enabled us to quickly develop the analysis workflow. We can hence report positive experience using \proglang{R} also for larger software components in production. 

The reporting component was realized using Microsoft Reporting Services~\citep{SSRS}, because this technology is widely used within the RKI.
It allows quick development of reports and works well with existing Microsoft Office tools, which the end-user, the epidemiologist, is used to. For example, one major requirement by the epidemiologists was to have the results compiled as Excel documents.
Moreover, pathogen-specific reports are automatically sent once a week by email to epidemiologists in charge of the respective pathogen.

Having state-of-the-art detection methods already implemented in \pkg{surveillance} helped us to focus on other challenges during development, such as bringing the system in the organization's workflow and finding ways to efficiently and effectively analyse about half a million of time series per day.
In addition, major developments in the \proglang{R} component can be shared with the community and are thus available to other public health institutes as well.

%% file: discussion.tex
\indent\indent  The \proglang{R} package \pkg{surveillance} was initially created as an implementational framework for the development and the evaluation of outbreak detection algorithms in
routine collected public health surveillance data. Throughout the years it has more and more also become a tool for the use of surveillance in routine practice. The presented description aimed at showing the potential of the package for aberration derection. Other functions offered by the package for modelling~\citep{meyer.etal2014} or back-projection of incidence cases~\citep{becker_marschner93} are documented elsewhere and contribute to widening the scope of possible analysis in infectious disease epidemiology when using \pkg{surveillance}. Future areas of interest for the package are \textit{e.g.}, to better take into account the multivariate and hierarchical structure of the data streams analysed. Another important topic is the adjustment for reporting delays when performing the surveillance. The package can be obtained from CRAN and resources for learning its use are listed in the documentation section of the project\footnote{\url{http://surveillance.r-forge.r-project.org/}~Online; accessed 17-February-2014}. As all \proglang{R} packages, \pkg{surveillance} is distributed with a manual describing each function with corresponding examples. The manual, the present article and two previous ones~\citep{hoehle2007, hoehle_mazick2009} form a good basis for getting started with the package. The data and analysis of the present manuscript are accessible as the vignette \texttt{"monitoringCounts.Rnw"} in the package.

Since all functionality is available just at the cost of learning \proglang{R} we hope that parts of the package can be useful in health facilities around the world. Even though the package is tailored for surveillance in public health contexts, properties such as overdispersion, low counts, presence of past outbreaks, apply to a wide range of count and categorical time series in other surveillance contexts such as financial surveillance~\citep{frisen2008financial}, occupational safety monitoring~\citep{accident} or environmental surveillance~\citep{Radio}.

Other \proglang{R} packages can be worth of interest to \pkg{surveillance} users. Statistical process control is offered by two other packages, \pkg{spc}~\citep{spc} and \pkg{qcc}~\citep{qcc}. The package \pkg{strucchange} allows detecting structural changes in general parametric models including GLMs~\citep{strucchange}. For epidemic modelling and outbreaks, packages such as \pkg{EpiEstim}~\citep{EpiEstim}, \pkg{outbreaker}~\citep{outbreaker} and \pkg{OutbreakTools}~\citep{OutbreakTools} offer good functionalities for investigating outbreaks that may for instance have been detected through to the use of \pkg{surveillance}. They are listed on the website of the \textit{R-epi project}\footnote{\url{https://sites.google.com/site/therepiproject}~Online; accessed 17-February-2014} that was initiated for compiling information about \proglang{R} tools useful for infectious diseases epidemiology. Another software of interest for aberration detection is SaTScan~\citep{SaTScan} which allows the detection of spatial, temporal and space-time clusters of events -- note that it is not a \proglang{R} package.

Code contributions to the package are very welcome as well as feedback and suggestions for improving the package.

%% file: acknowledgements.tex
The authors would like to express their gratitude to all contributors to the package, in particular 
Juliane Manitz, University of G\"{o}ttingen, Germany, for her work on the \texttt{boda} code and Angela Noufaily, The Open University, Milton Keynes, UK, for providing us the code used in her article that we extended for \texttt{farringtonFlexible}. We also thank Sebastian Meyer, University of Z\"{u}rich, Switzerland, for continuous technical support. The work of M. Salmon was financed by a PhD grant of the RKI.

%% file: master.bbl
\begin{thebibliography}{59}
\newcommand{\enquote}[1]{``#1''}
\providecommand{\natexlab}[1]{#1}
\providecommand{\url}[1]{\texttt{#1}}
\providecommand{\urlprefix}{URL }
\expandafter\ifx\csname urlstyle\endcsname\relax
  \providecommand{\doi}[1]{doi:\discretionary{}{}{}#1}\else
  \providecommand{\doi}{doi:\discretionary{}{}{}\begingroup
  \urlstyle{rm}\Url}\fi
\providecommand{\eprint}[2][]{\url{#2}}

\bibitem[{Bayer \emph{et~al.}(2014)Bayer, Bernard, Prager, Rabsch, Hiller,
  Malorny, Pfefferkorn, Frank, de~Jong, Friesema \emph{et~al.}}]{Newport2011}
Bayer C, Bernard H, Prager R, Rabsch W, Hiller P, Malorny B, Pfefferkorn B,
  Frank C, de~Jong A, Friesema I, \emph{et~al.} (2014).
\newblock \enquote{An Outbreak of Salmonella Newport Associated with Mung Bean
  Sprouts in Germany and the Netherlands, October to November 2011.}
\newblock \emph{Eurosurveillance}, \textbf{19}(1).

\bibitem[{{Becker} and {Marschner}(1993)}]{becker_marschner93}
{Becker} NG, {Marschner} IC (1993).
\newblock \enquote{A Method for Estimating the Age-Specific Relative Risk of
  {HIV} Infection from {AIDS} Incidence Data.}
\newblock \emph{Biometrika}, \textbf{80}(1).

\bibitem[{Bernard \emph{et~al.}(2014)Bernard, Werber, and
  Hohle}]{bernard_etal2014}
Bernard H, Werber D, Hohle M (2014).
\newblock \enquote{Estimating the under-reporting of norovirus illness in
  Germany utilizing enhanced awareness of diarrhoea during a large outbreak of
  Shiga toxin-producing E. coli O104:H4 in 2011 - a time series analysis.}
\newblock \emph{BMC Infectious Diseases}, \textbf{14}(1), 116.
\newblock \doi{10.1186/1471-2334-14-116}.

\bibitem[{Bivand \emph{et~al.}(2013)Bivand, Pebesma, and Gomez-Rubio}]{sp2}
Bivand RS, Pebesma E, Gomez-Rubio V (2013).
\newblock \emph{Applied Spatial Data Analysis With \proglang{R}, Second
  Edition}.
\newblock Springer-Verlag.

\bibitem[{{Brook} and {Evans}(1972)}]{brook_evans1972}
{Brook} D, {Evans} DA (1972).
\newblock \enquote{An Approach to the Probability Distribution of Cusum Run
  Length.}
\newblock \emph{Biometrika}, \textbf{59}(3), 539--549.

\bibitem[{Buckeridge \emph{et~al.}(2005)Buckeridge, Burkom, Campbell, Hogan,
  and Moore}]{Buckeridge2005}
Buckeridge DL, Burkom H, Campbell M, Hogan WR, Moore AW (2005).
\newblock \enquote{Algorithms for Rapid Outbreak Detection: a Research
  Synthesis.}
\newblock \emph{Journal of Biomedical Informatics}, \textbf{38}(2), 99 -- 113.
\newblock ISSN 1532-0464.
\newblock \doi{http://dx.doi.org/10.1016/j.jbi.2004.11.007}.

\bibitem[{Chen(1978)}]{Chen1978}
Chen R (1978).
\newblock \enquote{A Surveillance System for Congenital Malformations.}
\newblock \emph{Journal of the American Statistical Association},
  \textbf{73}(362), 323--327.
\newblock \doi{10.1080/01621459.1978.10481577}.

\bibitem[{Cori(2013)}]{EpiEstim}
Cori A (2013).
\newblock \emph{\pkg{EpiEstim}: a Package to Estimate Time Varying Reproduction
  Numbers from Epidemic Curves.}
\newblock \proglang{R} Package Version 1.1-2,
  \urlprefix\url{http://CRAN.R-project.org/package=EpiEstim}.

\bibitem[{Dowle \emph{et~al.}(2013)Dowle, Short, and Lianoglou}]{datatable2013}
Dowle M, Short T, Lianoglou S (2013).
\newblock \emph{\pkg{data.table}: Extension of \pkg{data.frame} for Fast
  Indexing, Fast Ordered Joins, Fast Assignment, Fast Grouping and List
  Columns.}
\newblock \proglang{R} Package Version 1.8.8,
  \urlprefix\url{http://CRAN.R-project.org/package=data.table}.

\bibitem[{Fahrmeir \emph{et~al.}(2013)Fahrmeir, Kneib, Lang, and
  Marx}]{fahrmeir2013regression}
Fahrmeir L, Kneib T, Lang S, Marx B (2013).
\newblock \emph{Regression: Models, Methods and Applications}.
\newblock Springer-Verlag.
\newblock ISBN 978-3-642-34333-9.

\bibitem[{{Farrington} and {Andrews}(2003)}]{farrington2003}
{Farrington} C, {Andrews} N (2003).
\newblock \enquote{Outbreak Detection: Application to Infectious Disease
  Surveillance.}
\newblock In R~{Brookmeyer}, D~{Stroup} (eds.), \emph{Monitoring the Health of
  Populations}, pp. 203--231. Oxford University Press.

\bibitem[{{Farrington} \emph{et~al.}(1996){Farrington}, {Andrews}, {Beale}, and
  Catchpole}]{farrington96}
{Farrington} C, {Andrews} N, {Beale} A, Catchpole M (1996).
\newblock \enquote{A Statistical Algorithm for the Early Detection of Outbreaks
  of Infectious Disease.}
\newblock \emph{Journal of the Royal Statistical Society A}, \textbf{159},
  547--563.

\bibitem[{Fricker \emph{et~al.}(2008)Fricker, Hegler, and Dunfee}]{SIM:SIM3197}
Fricker RD, Hegler BL, Dunfee DA (2008).
\newblock \enquote{Comparing Syndromic Surveillance Detection Methods: EARS'
  versus a CUSUM-Based Methodology.}
\newblock \emph{Statistics in Medicine}, \textbf{27}(17), 3407--3429.

\bibitem[{Fris{\'e}n(2008)}]{frisen2008financial}
Fris{\'e}n M (2008).
\newblock \emph{Financial surveillance}.
\newblock John Wiley \& Sons.

\bibitem[{Fris{\'e}n and Andersson(2009)}]{fri2009}
Fris{\'e}n M, Andersson E (2009).
\newblock \enquote{Semiparametric Surveillance of Monotonic Changes.}
\newblock \emph{Sequential Analysis}, \textbf{28}(4), 434--454.

\bibitem[{{Fris{\'e}n} \emph{et~al.}(2009){Fris{\'e}n}, {Andersson}, and
  {Schi{\"o}ler}}]{frisen_etal2009}
{Fris{\'e}n} M, {Andersson} E, {Schi{\"o}ler} L (2009).
\newblock \enquote{Robust Outbreak Surveillance of Epidemics in Sweden.}
\newblock \emph{Statistics in Medicine}, \textbf{28}, 476--493.

\bibitem[{{Held} \emph{et~al.}(2006){Held}, {Hofmann}, {H{\"o}hle}, and
  {Schmid}}]{held_etal2006}
{Held} L, {Hofmann} M, {H{\"o}hle} M, {Schmid} V (2006).
\newblock \enquote{A Two Component Model for Counts of Infectious Diseases.}
\newblock \emph{Biostatistics}, \textbf{7}, 422--437.

\bibitem[{{Held} \emph{et~al.}(2005){Held}, {H{\"o}hle}, and
  {Hofmann}}]{held_etal2005}
{Held} L, {H{\"o}hle} M, {Hofmann} M (2005).
\newblock \enquote{{A Statistical Framework for the Analysis of Multivariate
  Infectious Disease Surveillance Data}.}
\newblock \emph{Statistical Modelling}, \textbf{5}, 187--199.

\bibitem[{{H{\"o}hle}(2007)}]{hoehle2007}
{H{\"o}hle} M (2007).
\newblock \enquote{\pkg{surveillance}: An \proglang{R} Package for the
  Monitoring of Infectious Diseases.}
\newblock \emph{Computational Statistics}, \textbf{22}(4), 571--582.

\bibitem[{H\"{o}hle(2010)}]{hoehle2010}
H\"{o}hle M (2010).
\newblock \enquote{Online Change-Point Detection in Categorical Time Series.}
\newblock In T~Kneib, G~Tutz (eds.), \emph{Statistical Modelling and Regression
  Structures}, pp. 377--397. Physica-Verlag HD.
\newblock \doi{10.1007/978-3-7908-2413-1_20}.

\bibitem[{{H{\"o}hle} and {Mazick}(2010)}]{hoehle_mazick2009}
{H{\"o}hle} M, {Mazick} A (2010).
\newblock \enquote{Aberration Detection in \proglang{R} Illustrated by {D}anish
  Mortality Monitoring.}
\newblock In T~{Kass-Hout}, X~{Zhang} (eds.), \emph{Biosurveillance: A Health
  Protection Priority}, chapter~12, pp. 215--238. CRC Press.

\bibitem[{{H{\"o}hle} and {Paul}(2008)}]{hoehle_paul2008}
{H{\"o}hle} M, {Paul} M (2008).
\newblock \enquote{Count Data Regression Charts for the Monitoring of
  Surveillance Time Series.}
\newblock \emph{Computational Statistics \& Data Analysis}, \textbf{52}(9),
  4357--4368.

\bibitem[{{Hulth} \emph{et~al.}(2010){Hulth}, {Andrews}, {Ethelberg},
  {Dreesman}, {Faensen}, {van Pelt}, and {Schnitzler}}]{hulth_etal2010}
{Hulth} A, {Andrews} N, {Ethelberg} S, {Dreesman} J, {Faensen} D, {van Pelt} W,
  {Schnitzler} J (2010).
\newblock \enquote{Practical Usage of Computer-Supported Outbreak Detection in
  Five European Countries.}
\newblock \emph{Eurosurveillance}, \textbf{15}(36).

\bibitem[{Jombart \emph{et~al.}(2014)Jombart, Cori, Didelot, Cauchemez, Fraser,
  and Ferguson}]{outbreaker}
Jombart T, Cori A, Didelot X, Cauchemez S, Fraser C, Ferguson N (2014).
\newblock \enquote{Bayesian Reconstruction of Disease Outbreaks by Combining
  Epidemiologic and Genomic Data.}
\newblock \emph{PLoS Comput Biol}, \textbf{10}(1), e1003457.
\newblock \doi{10.1371/journal.pcbi.1003457}.

\bibitem[{Knoth(2014)}]{spc}
Knoth S (2014).
\newblock \emph{\pkg{spc}: Statistical Process Control -- Collection of Some
  Useful Functions}.
\newblock \proglang{R} Package Version 0.5.0,
  \urlprefix\url{http://CRAN.R-project.org/package=spc}.

\bibitem[{Kulldorff(1997)}]{SaTScan}
Kulldorff M (1997).
\newblock \emph{\proglang{SaTScan}: Software for the Spatial, Temporal and
  Space-Time Scan Statistics}.
\newblock Boston, MA, USA.
\newblock \urlprefix\url{http://www.satscan.org/}.

\bibitem[{Lawless(1987)}]{lawless1987}
Lawless JF (1987).
\newblock \enquote{Negative Binomial and Mixed Poisson Regression.}
\newblock \emph{Canadian Journal of Statistics}, \textbf{15}(3), 209--225.

\bibitem[{Lucas and Crosier(1982)}]{lucas1982fast}
Lucas JM, Crosier RB (1982).
\newblock \enquote{Fast Initial Response for CUSUM Quality-Control Schemes:
  Give Your CUSUM a Head Start.}
\newblock \emph{Technometrics}, \textbf{24}(3), 199--205.

\bibitem[{Luo \emph{et~al.}(2012)Luo, DeVol, and Sharp}]{Radio}
Luo P, DeVol TA, Sharp JL (2012).
\newblock \enquote{CUSUM Analyses of Time-Interval Data for Online Radiation
  Monitoring.}
\newblock \emph{Health physics}, \textbf{102}(6), 637--645.

\bibitem[{Manitz and H\"{o}hle(2013)}]{Manitz2013}
Manitz J, H\"{o}hle M (2013).
\newblock \enquote{Bayesian Outbreak Detection Algorithm for Monitoring
  Reported Cases of Campylobacteriosis in Germany.}
\newblock \emph{Biometrical Journal}, \textbf{55}(4), 509--526.
\newblock ISSN 1521-4036.
\newblock \doi{10.1002/bimj.201200141}.

\bibitem[{{Meyer} \emph{et~al.}(2012){Meyer}, {Elias}, and
  {H\"{o}hle}}]{meyer_etal2012}
{Meyer} S, {Elias} J, {H\"{o}hle} M (2012).
\newblock \enquote{A Space-Time Conditional Intensity Model for Invasive
  Meningococcal Disease Occurrence.}
\newblock \emph{Biometrics}, \textbf{68}(2), 607--616.

\bibitem[{Meyer \emph{et~al.}(2014)Meyer, Held, and H\"{o}hle}]{meyer.etal2014}
Meyer S, Held L, H\"{o}hle M (2014).
\newblock \enquote{{S}patio-Temporal Analysis of Epidemic Phenomena Using the
  \proglang{R} Package \pkg{surveillance}.}
\newblock In preparation.

\bibitem[{{Microsoft Corp.}(2012{\natexlab{a}})}]{SSAS}
{Microsoft Corp} (2012{\natexlab{a}}).
\newblock \emph{Microsoft SQL Server Analysis Services, Version~2012}.
\newblock \urlprefix\url{http://www.microsoft.com/}.

\bibitem[{{Microsoft Corp.}(2012{\natexlab{b}})}]{SSRS}
{Microsoft Corp} (2012{\natexlab{b}}).
\newblock \emph{Microsoft SQL Server Reporting Services, Version~2012}.
\newblock \urlprefix\url{http://www.microsoft.com/}.

\bibitem[{{Noufaily} \emph{et~al.}(2012){Noufaily}, {Enki}, {Farrington},
  {Garthwaite}, {Andrews}, and {Charlett}}]{Noufaily2012}
{Noufaily} A, {Enki} D, {Farrington} P, {Garthwaite} P, {Andrews} N, {Charlett}
  A (2012).
\newblock \enquote{An Improved Algorithm for Outbreak Detection in Multiple
  Surveillance Systems.}
\newblock \emph{Statistics in Medicine}.

\bibitem[{Pebesma and Bivand(2005)}]{sp1}
Pebesma EJ, Bivand RS (2005).
\newblock \enquote{Classes and Methods for Spatial Data In \proglang{R}.}
\newblock \emph{R News}, \textbf{5}(2), 9--13.
\newblock \urlprefix\url{http://CRAN.R-project.org/doc/Rnews/}.

\bibitem[{{Pierce} and {Schafer}(1986)}]{pierce_schafer86}
{Pierce} D, {Schafer} D (1986).
\newblock \enquote{Residuals in Generalized Linear Models.}
\newblock \emph{Journal of the American Statistical Association},
  \textbf{81}(396), 977--986.

\bibitem[{Reynolds and Stoumbos(2000)}]{Reynolds2000}
Reynolds M, Stoumbos Z (2000).
\newblock \enquote{A general approach to modeling CUSUM charts for a
  proportion.}
\newblock \emph{IIE Transactions}, \textbf{32}(6), 515--535.

\bibitem[{{Riebler}(2004)}]{riebler2004}
{Riebler} A (2004).
\newblock \emph{{Empirischer Vergleich von statistischen Methoden zur
  Ausbruchserkennung bei Surveillance Daten}}.
\newblock Master's thesis, Department of Statistics, University of Munich.
\newblock Bachelor's thesis.

\bibitem[{Rigby and Stasinopoulos(2005)}]{Rigby2005}
Rigby RA, Stasinopoulos DM (2005).
\newblock \enquote{Generalized Additive Models for Location, Scale and Shape.}
\newblock \emph{Journal of the Royal Statistical Society C}, \textbf{54}(3),
  507--554.
\newblock \doi{10.1111/j.1467-9876.2005.00510.x}.

\bibitem[{Ripley and Lapsley(2012)}]{rodbc2013}
Ripley B, Lapsley M (2012).
\newblock \emph{\pkg{RODBC}: ODBC Database Access}.
\newblock \proglang{R} Package Version 1.3-6,
  \urlprefix\url{http://CRAN.R-project.org/package=RODBC}.

\bibitem[{{Rogerson} and {Yamada}(2004)}]{rogerson_yamada2004}
{Rogerson} P, {Yamada} I (2004).
\newblock \enquote{Approaches to Syndromic Surveillance When Data Consist of
  Small Regional Counts.}
\newblock \emph{Morbidity and Mortality Weekly Report}, \textbf{53}, 79--85.

\bibitem[{{Rossi} \emph{et~al.}(1999){Rossi}, {Lampugnani}, and
  {Marchi}}]{rossi_etal99}
{Rossi} G, {Lampugnani} L, {Marchi} M (1999).
\newblock \enquote{An Approximate {CUSUM} Procedure for Surveillance of Health
  Events.}
\newblock \emph{Statistics in Medicine}, \textbf{18}, 2111--2122.

\bibitem[{{Rue} \emph{et~al.}(2013){Rue}, {Martino}, {Lindgren}, {Simpson}, and
  {Riebler}}]{INLA}
{Rue} H, {Martino} S, {Lindgren} F, {Simpson} D, {Riebler} A (2013).
\newblock \emph{\pkg{INLA}: Functions Which Allow to Perform Full Bayesian
  Analysis of Latent {G}aussian Models Using {I}ntegrated {N}ested {L}aplace
  {A}pproximation}.
\newblock \proglang{R} Package Version 0.0-1386250221.

\bibitem[{Ryan and Ulrich(2014)}]{xts}
Ryan JA, Ulrich JM (2014).
\newblock \emph{Package \pkg{xts}}.
\newblock \proglang{R} Package Version 0.9-7,
  \urlprefix\url{http://CRAN.R-project.org/package=xts}.

\bibitem[{Schuh \emph{et~al.}(2013)Schuh, Camelio, and Woodall}]{accident}
Schuh A, Camelio JA, Woodall WH (2013).
\newblock \enquote{Control Charts for Accident Frequency: a Motivation for
  Real-Time Occupational Safety Monitoring.}
\newblock \emph{International Journal of Injury Control and Safety Promotion},
  pp. 1--9.
\newblock \doi{10.1080/17457300.2013.792285}.

\bibitem[{{Schumacher} \emph{et~al.}(2014){Schumacher}, {Salmon}, {Frank},
  {Claus}, and H{\"o}hle}]{Dirk}
{Schumacher} D, {Salmon} M, {Frank} C, {Claus} H, H{\"o}hle M (2014).
\newblock \enquote{Automated outbreak detection system for notifiable diseases
  in {G}ermany.}
\newblock Submitted.

\bibitem[{Scrucca(2004)}]{qcc}
Scrucca L (2004).
\newblock \enquote{\pkg{qcc}: an \proglang{R} Package for Quality Control
  Charting and Statistical Process Control.}
\newblock \emph{R News}, \textbf{4/1}, 11--17.
\newblock \urlprefix\url{http://CRAN.R-project.org/doc/Rnews/}.

\bibitem[{Shmueli and Burkom(2010)}]{Shmueli2010}
Shmueli G, Burkom H (2010).
\newblock \enquote{Statistical Challenges Facing Early Outbreak Detection in
  Biosurveillance.}
\newblock \emph{Technometrics}, \textbf{52}(1), 39--51.
\newblock \doi{10.1198/TECH.2010.06134}.

\bibitem[{Sonesson and Bock(2003)}]{Sonesson2003}
Sonesson C, Bock D (2003).
\newblock \enquote{A Review and Discussion of Prospective Statistical
  Surveillance in Public Health.}
\newblock \emph{Journal of the Royal Statistical Society A}, \textbf{166}(1),
  5--21.
\newblock ISSN 1467-985X.

\bibitem[{Stasinopoulos and Rigby(2007)}]{StasJSS}
Stasinopoulos DM, Rigby RA (2007).
\newblock \enquote{Generalized Additive Models for Location Scale and Shape
  (GAMLSS) in \proglang{R}.}
\newblock \emph{Journal of Statistical Software}, \textbf{23}(7), 1--46.
\newblock ISSN 1548-7660.
\newblock \urlprefix\url{http://www.jstatsoft.org/v23/i07}.

\bibitem[{{Steiner} \emph{et~al.}(1999){Steiner}, {Cook}, and
  {Farewell}}]{Steiner1999}
{Steiner} SH, {Cook} RJ, {Farewell} VT (1999).
\newblock \enquote{Monitoring Paired Binary Surgical Outcomes Using Cumulative
  Sum Charts.}
\newblock \emph{Statistics in Medicine}, \textbf{18}, 69--86.

\bibitem[{{Stroup} \emph{et~al.}(1989){Stroup}, {Williamson}, {Herndon}, and
  {Karon}}]{stroup_etal89}
{Stroup} D, {Williamson} G, {Herndon} J, {Karon} J (1989).
\newblock \enquote{Detection of Aberrations in the Occurrence of Notifiable
  Diseases Surveillance Data.}
\newblock \emph{Statistics in Medicine}, \textbf{8}, 323--329.

\bibitem[{{The Hackout team}(2014)}]{OutbreakTools}
{The Hackout team} (2014).
\newblock \emph{\pkg{OutbreakTools}: Basic Tools for the Analysis of Disease
  Outbreaks.}
\newblock \proglang{R} Package Version 0.1-0,
  \urlprefix\url{http://CRAN.R-project.org/package=OutbreakTools}.

\bibitem[{Unkel \emph{et~al.}(2012)Unkel, Farrington, Garthwaite, Robertson,
  and Andrews}]{Unkel2012}
Unkel S, Farrington CP, Garthwaite PH, Robertson C, Andrews N (2012).
\newblock \enquote{Statistical Methods for the Prospective Detection of
  Infectious Disease Outbreaks: a Review.}
\newblock \emph{Journal of the Royal Statistical Society A}, \textbf{175}(1),
  49--82.
\newblock ISSN 1467-985X.
\newblock \doi{10.1111/j.1467-985X.2011.00714.x}.

\bibitem[{Wickham(2013)}]{testthat2013}
Wickham H (2013).
\newblock \emph{\pkg{testthat}: Testthat Code. Tools to Make Testing Fun :)}.
\newblock \proglang{R} Package Version 0.7.1,
  \urlprefix\url{http://CRAN.R-project.org/package=testthat}.

\bibitem[{Zeileis and Grothendieck(2005)}]{zoo}
Zeileis A, Grothendieck G (2005).
\newblock \enquote{zoo: S3 Infrastructure for Regular and Irregular Time
  Series.}
\newblock \emph{Journal of Statistical Software}, \textbf{14}(6), 1--27.
\newblock \urlprefix\url{http://www.jstatsoft.org/v14/i06/}.

\bibitem[{Zeileis \emph{et~al.}(2002)Zeileis, Leisch, Hornik, and
  Kleiber}]{strucchange}
Zeileis A, Leisch F, Hornik K, Kleiber C (2002).
\newblock \enquote{\pkg{strucchange}: An \proglang{R} Package for Testing for
  Structural Change in Linear Regression Models.}
\newblock \emph{Journal of Statistical Software}, \textbf{7}(2), 1--38.
\newblock ISSN 1548-7660.
\newblock \urlprefix\url{http://www.jstatsoft.org/v07/i02}.

\bibitem[{Zhang and Zhou(2014)}]{MGLM}
Zhang Y, Zhou H (2014).
\newblock \emph{\pkg{MGLM}: Multivariate Response Generalized Linear Models}.
\newblock \proglang{R} Package Version 0.0.4,
  \urlprefix\url{http://CRAN.R-project.org/package=MGLM}.

\end{thebibliography}
